\documentclass[12pt, reqno]{report}

\makeatletter
\g@addto@macro{\endabstract}{\@setabstract}
\makeatother

\usepackage{graphics, stackrel}
\usepackage{amsmath, amssymb, amsthm}
\usepackage{graphicx}
\usepackage{verbatim}
\usepackage{amsfonts}
\usepackage{xcolor}
\usepackage{lipsum}
\usepackage{float}
\usepackage{bbm}
\usepackage{titling}
\usepackage{blindtext}
\usepackage{natbib}
\usepackage{tocbibind}
\usepackage{hyperref}
\hypersetup{citecolor=blue, colorlinks=true, linkcolor=blue}

\usepackage{mathpazo}
\usepackage{lmodern}
\usepackage{classicthesis}
\usepackage[T1]{fontenc}

\usepackage{booktabs}
\usepackage{fancyvrb}
\usepackage{color}
\usepackage{mdwlist}
\usepackage{mathtools}
\usepackage{accents}
\usepackage[normalem]{ulem}

\usepackage{enumitem}
\setlist[enumerate]{itemsep=2pt,topsep=3pt}
\setlist[itemize]{itemsep=2pt,topsep=3pt}
\setlist[enumerate,1]{label=(\alph*)}

\usepackage{caption}
\usepackage{subcaption}

\usepackage{graphicx}
\usepackage{amsmath}
\makeatletter
\newcommand{\distas}[1]{\mathbin{\overset{#1}{\kern\z@\sim}}}%
\newsavebox{\mybox}\newsavebox{\mysim}
\newcommand{\distras}[1]{%
  \savebox{\mybox}{\hbox{\kern3pt$\scriptstyle#1$\kern3pt}}%
  \savebox{\mysim}{\hbox{$\sim$}}%
  \mathbin{\overset{#1}{\kern\z@\resizebox{\wd\mybox}{\ht\mysim}{$\sim$}}}%
}
\makeatother

\usepackage{mathrsfs}
\usepackage{bbm}

\usepackage[left=1.1in, right=1.1in, top=0.85in, bottom=1.05in, includehead, includefoot]{geometry}

\renewcommand{\leq}{\leqslant}
\renewcommand{\geq}{\geqslant}

\newcommand{\HRule}{\rule{\linewidth}{0.3mm}}

\theoremstyle{plain}
\newtheorem{theorem}{Theorem}[section]
\newtheorem{corollary}[theorem]{Corollary}
\newtheorem{lemma}[theorem]{Lemma}
\newtheorem{proposition}[theorem]{Proposition}

\theoremstyle{definition}
\newtheorem{definition}{Definition}[section]

\newtheorem{remark}{Remark}[section]
\newtheorem{assumption}{Assumption}[section]

\usepackage{xr-hyper}

\newcommand{\RNum}[1]{\uppercase\expandafter{\romannumeral #1\relax}}
\def\ifNumPutRoman#1{%
  \if!\ifnum9<1#1!\else_\fi
    \RNum{#1}\else#1\fi}

\titleformat{name=\chapter}[display]%
        {\relax}
        {\centering{ \ifNumPutRoman{\thechapter}} \\ }
        {10pt}%
        {\centering\spacedallcaps}[\normalsize\vspace*{2\baselineskip}]%

\numberwithin{equation}{section}
\numberwithin{theorem}{section}

\begin{document}
\linespread{1.3}

\title{ }
\date{}
\author{}
\begin{titlepage} 
	
	\center 
	
\begin{figure}[H]
    \centering
    \includegraphics[scale=0.9]{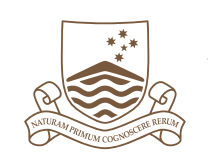}
\end{figure}

	\textsc{\LARGE The Australian National University}\\[1.5cm] 
	
	\textsc{\Large Research School of Economics}\\[0.5cm] 
	
	\textsc{\large Honours Thesis}\\[0.5cm] 
	
	
	\HRule\\[0.4cm]
	
	{\LARGE\bfseries Existence and Uniqueness of Recursive Utility Models in $L_p$ }\\[0.4cm] 
	
	\HRule\\[1.5cm]
	
	
	\begin{minipage}{0.4\textwidth}
		\begin{flushleft}
			\large
			\textit{Author}\\
			Flint \textsc{O'Neil} 
		\end{flushleft}
	\end{minipage}
	~
	\begin{minipage}{0.4\textwidth}
		\begin{flushright}
			\large
			\textit{Supervisor}\\
			Prof. John \textsc{Stachurski} 
		\end{flushright}
	\end{minipage}
	
	\vfill
	
	\hfill
	
	\hfill
	
 \begin{center}
    A Thesis Submitted in Partial Fulfilment\\
    of the Requirements for the Degree of\\
    Bachelor of Economics (Honours)
    
    \hfill
    
  \end{center}
	
	
	\vfill\vfill\vfill 
	
	{\large\today} 
	
	
	 
	
	\vfill 
	
\end{titlepage}








    
    

    
     
    




\chapter*{Declaration}
This thesis contains no material which has been accepted for the award of any other degree
or diploma in any University, and, to the best of my knowledge and belief, contains no
material published or written by another person, except where due reference is made in the
thesis.
 
\vspace{20mm}  

\hspace{80mm}\rule{40mm}{.15mm}\par   
\hspace{80mm} Flint O'Neil \par
\hspace{80mm} \today

\chapter*{Acknowledgements}
 *Redacted*
 
\vspace{20mm}  

\hspace{80mm}\rule{40mm}{.15mm}\par   

\chapter*{Abstract}
 Recursive preferences, of the sort developed by \cite{EpsteinZin1989}, play an integral role in modern macroeconomics and asset pricing theory. 
Unfortunately, it is non-trivial to establish the unique existence of a solution to recursive utility models. We show that the tightest known existence and uniqueness conditions can be extended to (i) \cite{Schorfheide} recursive utilities and (ii) recursive utilities with `narrow framing'. Further, we sharpen the solution space of \cite{BorovickaStachurski2017} from $L_1$ to $L_p$ so that the results apply to a broader class of modern asset pricing models. For example, using $L_2$ Hilbert space theory, we find the class of parameters which generate a unique $L_2$ solution to the \cite{bansal1} and \cite{Schorfheide} models.
 
\vspace{20mm}  

\hspace{80mm}\rule{40mm}{.15mm}\par

\clearpage

{\hypersetup{linkcolor=black}
\tableofcontents
}

\clearpage

\chapter{Introduction}
\textit{``If the theory disagrees with the data, you throw out the data'' - Rabee Tourky}

Economic agents often make decisions under uncertainty. Whether it be firms calculating optimal inventories or hedge funds seeking to maximise a portfolio's value, choice is inextricably linked to risk. Generally, economists analyse risk by assuming agents have additively separable preferences over states of nature. As a result, an agent's risk aversion cannot be disentangled from their elasticity of substitution. This inability to isolate risk preferences has consistently challenged economic models. In finance for example, \cite{MehraPrescott}, \cite{weil1990} and \cite{bansal1} note that standard economic theories of risk struggle to account for the 6\% equity risk premium.

This motivates our interest in the \cite{EpsteinZin1989} class of recursive utility functions. Whereas additively separable preferences combine risk aversion and intertemporal substitution, Epstein-Zin recursive preferences disaggregate these two forces. This distinction means that Epstein-Zin recursive preferences are particularly powerful in asset pricing and long-run risk models. Indeed, Roger Farmer notes that in the context of finance theory, ``the dominant view [...] is that people maximize the [...] value of [...] preferences first formalized by Epstein and Zin.''\footnote{ These comments can be found on Roger Farmer's personal blog at:

http://rogerfarmerblog.blogspot.com/2015/07/behavioural-economics-and-exotic.html}

The dominance of Epstein-Zin recursive utilites in finance stems from their success in explaining empirical asset pricing facts. For example, \cite{bansal1} puts forward an economic mechanism of long-run risk which relies on an Epstein-Zin specification. \cite{pohl2} describes this to be the ``foundation for a large literature on the ability of long-run risk to solve empirical puzzles''.\footnote{Comments found in \cite{pohl2}}

Simultaneously, \cite{bansal1} apply the \cite{CampbellShiller} log-linearisation technique to analyse recursive utilities. But \cite{pohl2} show that long-run risk models exhibit economically significant non-linearities. Thus, log-linearisation introduces large numerical errors. This highlights the necessity of understanding the full Epstein-Zin recursive utility model, not just a linear approximation.





Unfortunately, it is difficult to establish the existence of a unique solution to Epstein-Zin recursive utilities. This poses a severe limitation; without a unique solution, utility representations lack any informative content. \cite{BorovickaStachurski2017} provide existence and uniqueness conditions which are both necessary \textit{and} sufficient for classic Epstein-Zin recursive utilities. The conditions are ``as tight as possible in a range of empirically plausible settings''\footnote{ Comment found in \cite{BorovickaStachurski2017} },  though can be hard to evaluate analytically. This paper aims to build upon the results derived by \cite{BorovickaStachurski2017}.

This paper's central results are theorems \ref{BIG}, \ref{unbounded}, and proposition \ref{narrow}. These theorems provide conditions for existence and uniqueness of recursive utilities. The closest existing result in the literature is \cite{BorovickaStachurski2017}, theorem 3.1. The theorems in this thesis extend the \cite{BorovickaStachurski2017} result in two significant ways.

First, this paper's results consider newer classes of recursive utility. Specifically, we consider recursive utilities with (i) `time preference' shocks and (ii) narrow framing. The latter consideration has troubled economists for some time, with \cite{guo} having only been able to establish existence on a finite state space. Our results thus provide mathematical foundation to recent asset pricing papers, including \cite{Albuquerquemain}, \cite{Schorfheide} and \cite{barberis2006}.


Second, our results generalise the solution space from $L_1(\mathbb{X})$ to $L_p(\mathbb{X})$. This generalisation is important in both an applied and theoretical sense. Regarding application, $L_1$ solutions can be of limited practical use due to infinite second moments. When one solves for the utility value of a consumption stream, the result is a `price' over the stream. This price is typically a function of the current (Markovian) state. If the resulting price function has infinite variance, then it is of limited use to econometricians. By extending to higher $L_p$, we guarantee finite moments and thus provide conditions which are more useful to empirical analysis.

Regarding theory, \cite{mandelbrot} and others identify that asset pricing data is typically heavy tailed. This means that it is important to consider unbounded state spaces, as bounded approximations can be misinformative. By widening the solution space to $L_p$, our results apply to a larger set of unbounded asset pricing specifications. This is because compactness conditions, needed for regularity, are more readily met in higher $L_p$. To illustrate, this paper tackles (partially) unbounded specifications of the \cite{bansal1} and \cite{Schorfheide} models in $L_2$. These are major long-run risk papers, and existence of a unique solution has been an active question. Our results establishing existence and uniqueness here are thus a major development.

This paper is structured as follows. Chapter 2 canvasses recent developments in the recursive utility literature. Chapter 3 presents the main findings. Chapter 4 applies these findings to the major long-run risk models. Chapter 5 is dedicated to discussing results. The appendix contains the vast majority of mathematical proofs. Although this paper is primarily theoretical, it is theoretical with a view towards results rather than technique.

\clearpage

\chapter{Literature Review}

\textit{``You only talk about \cite{bansal1}. Why don't you ever ask me about my day?'' - Laksshini Sundaramoorthy}

\section{Limitations to von Neumann-Morgenstern Preferences}

The \cite{VNM} (vNM) expected utility representation is the workhorse model of decision-making under uncertainty. Time is indexed discretely by $t \in \mathbb{N}$ and streams of risk-contingent consumption are ranked according to
\begin{equation}\label{VNM1} V_t = \mathbb{E}_t \Big[\sum^\infty_{t=0} \beta^t u(c_t)\Big] \end{equation}
where $\beta \in (0,1)$, and $u(\cdot)$ is the one-period utility function. 

Under standard assumptions on $u(\cdot) \geq 0$ (see, for example \cite{LucasStokey}), the sequence of partial sums converges from below to the infinite horizon specification. That is,
$$V_t^n = \sum^n_{t=0} \beta^t u(c_t) \nearrow \sum^\infty_{t=0} \beta^t u(c_t) = V_t.$$ 
Applying the monotone convergence theorem, we may rewrite \eqref{VNM1} as $$ V_t = \sum^\infty_{t=0} \mathbb{E}_t \Big[ \beta^t u(c_t)\Big].$$
This last expression is more useful as it can be written recursively as
\begin{equation}\label{rec}
V_t = u(c_t) + \beta \mathbb{E}(V_{t+1}).\footnote{In this recursion an agent's optimal decisions will be dynamically consistent. This is an important property for most economic models.}
\end{equation}

Although this utility representation is both simple and powerful, it is not without limitations. One significant drawback is that the vNM representation does not disentangle risk aversion from preferences over intertemporal substitution. Accordingly, vNM utilities have not found total empirical success in financial economics and macroeconomics (e.g see \cite{MehraPrescott} and \cite{HansenSingleton}).

\section{ Models of Recursive Preferences }



Recursive preferences, pioneered by \cite{KrepsPorteus1978}, \cite{EpsteinZin1989} and \cite{weil1990}, generalise equation \eqref{rec}. In the \eqref{rec} recursion, present value is a function of consumption `today' and value `tomorrow'. Drawing from this notion, \cite{EpsteinZin1989} define the class of stationary recursive preferences as
$$V_t = W[c_t, f(V_{t+1})].$$
This specification consists of two main components: a time aggregator representing time preference, $W$, and a `Kreps-Porteus' certainty equivalent capturing risk aversion, $f$. To distinguish intertemporal substitution from risk-aversion, Epstein-Zin utilities allow for preference over the \textit{timing} of the resolution of uncertainty.\footnote{As noted by \cite{backus}, preferences derived this way are stationary and dynamically consistent.}


Epstein-Zin utilities are integral to modern financial economics, particularly long-run risk models. \cite{bansal1} use Epstein-Zin to generate time-varying risk premia to justify the `excess volatility' of asset prices identified in \cite{shiller}. Subsequent long-run risk contributions relying on Epstein-Zin utilities include \cite{HansenHeaton}, \cite{BansalKikuYaron2012}, \cite{bansal-kiku-shaliastovich-yaron2014} and \cite{Schorfheide} among others.

Epstein-Zin utilities are also fundamental to macroeconomics. For instance, \cite{tallarini} and \cite{dolmas} examine the welfare effects of business cycles on agents with Epstein-Zin utility. Both papers illustrate how models incorporating vNM utility underestimate the welfare costs of macroeconomic volatility.

Nevertheless, recursive utility is not without its flaws. As \cite{CampbellAmmer1993} and \cite{Cochrane2011} note, variation in asset returns is overwhelmingly due to variation in discount factors. Although recursive utilities are able to isolate risk aversion, they do not consider time-varying discount factors. To address this shortcoming, \cite{Albuquerquemain} add `time-preference' discount shocks to the recursive utility valuation. This augmentation has seen some success. For example, the long-run risk model of \cite{Schorfheide} uses `time preference' shocks to estimate asset price persistence.

\section{ Existence and Uniqueness }

In recursive utility models, a consumption stream's value is found by solving a nonlinear, forward-looking difference equation. It is thus non-trivial to establish the existence of a unique solution. Originally, sufficient conditions were provided by \cite{EpsteinZin1989}, and then built upon by \cite{marinacci} and \cite{pohl}. However, the proposed conditions require the asymptotic consumption rate, $\frac{C_{t+1}}{C_t}$, to be \textit{almost surely} bounded. This in turn renders recursive utilities inapplicable to most asset pricing models.

To achieve a tighter result, \cite{BorovickaStachurski2017} exploit a link between recursive utilities and a Perron-Frobenius eigenvalue problem. The authors show that a unique solution can be found by considering the \textit{average} ``across all paths''\footnote{This comment is found in \cite{BorovickaStachurski2017}}. This condition is much weaker than requiring uniform bounds on the upper tail of the distribution.

 \cite{BorovickaStachurski2017} build upon \cite{HansenScheinkman}. Although \cite{HansenScheinkman} treat unbounded consumption paths, they only show the existence of a solution for some preference parameters. By contrast, \cite{BorovickaStachurski2017} allow for all parameters, while also establishing sufficient \textit{and} necessary conditions.


This paper extends \cite{BorovickaStachurski2017}, to make two contributions. First, we establish parallel results for recursive utilities with (i) time preference shocks and (ii) narrow framing. The latter consideration in particular has confounded economists for some time, with \cite{guo} having only been able to establish existence on a finite state space. Second, we generalise the solution space from 
 $L_1(\mathbb{X})$ to $L_p(\mathbb{X})$. As mentioned earlier, this allows for (i) solutions with greater empirical content and (ii) broader applications in an unbounded state space. Indeed, this thesis proves new results for \cite{bansal1} and \cite{Schorfheide}.

\clearpage

\chapter{Recursive Utility Models: Existence, Uniqueness and Stability}

\textit{``One day you'll realise that theory should have empirical content.'' - Tim Kam}

\section{Setup and Intuition}

The \cite{EpsteinZin1989} model of recursive utility defines preferences by
\begin{equation}
    V_t = \Big[ (1 - \beta) C_t^{1-1/\psi} + \beta \{ \mathcal{R}_t(V_{t+1}) \}^{1-1/\psi} \Big]^{1/(1-1/\psi)}
\end{equation}
where $\beta \in (0,1)$ is a time discount factor, $\{C_t\}$ is a consumption path and $V_t$ is the utility value of the path extending from time $t$. The scalar $\psi \neq 1$ captures the elasticity of intertemporal substitution (IES). The function $\mathcal{R}_t$ is the Kreps-Porteus certainty equivalent defined by
\begin{equation}
    \mathcal{R}_t(V_{t+1}) = (\mathbb{E}_t V_{t+1}^{1 - \gamma})^{1 /(1 - \gamma)}.
\end{equation}
where $\gamma \neq 1$ uniquely governs risk aversion. The insight of \cite{KrepsPorteus1978} is that this equation imposes a preference over the \textit{timing of the resolution of uncertainty} (TRU). This breaks the link between risk aversion and IES.

The function $\mathcal{R}_t$ is a certainty equivalent of future utility because \cite{EpsteinZin1989} implicitly take $u(c_t) = c_t^{1-1/\psi}$. The certainty equivalent is over \textit{utilities}, not \textit{consumption}, as a result of preference over the TRU. In the Epstein-Zin representation, the agent chooses to trade-off between utility `today' and a certainty equivalent of value `tomorrow'.



In \cite{Schorfheide}, the Epstein-Zin recursion is modified to include a `time preference shock', $\lambda_t$, so that lifetime value takes the form
\begin{equation}\label{eppy}
    V_t = \Big[ (1 - \beta) \lambda_t C_t^{1-1/\psi} + \beta \{ \mathcal{R}_t(V_{t+1}) \}^{1-1/\psi} \Big]^{1/(1-1/\psi)}
\end{equation}
The shocks $\{\lambda_t\}_{t=0}^\infty$ are a function of the state process. They are restricted to attain values in a compact set.

In this paper, we surmise a general Markov environment on a state space $\mathbb{X}$. This setting involves two mathematical assumptions.

\begin{assumption}\label{ass1}
Consumption growth is specified according to
\begin{equation}
    \ln(C_{t+1}/C_t ) = \kappa(X_t, X_{t+1},\epsilon_{t+1})
\end{equation}
where $\kappa$ is continuous and $\{X_t\} \subset \mathbb{X}$ is the exogenous Markov state process. The innovation process $\{\epsilon_t\}$ is IID on $\mathbb{R}^k$, independent of $\{X_t\}$.

We let $p(x,\cdot)$ represent the stochastic transition kernel for $X_{t+1}$ given that $X_t = x$. Assumption \ref{ass1} is standard in the literature: see \cite{HansenScheinkman}, \cite{BorovickaStachurski2017} and \cite{guo}.
\end{assumption}

\begin{assumption}\label{ass2}
The transition kernel $p$ is jointly continuous in its arguments. Moreover, for some $\ell > 0$, $p^\ell$ is everywhere positive. This `irreducibility' assumption ensures ergodicity, and hence $\{X_t\}$ converges to a unique stationary distribution, which we denote by $\pi$.
\end{assumption}

Let $\mathcal{F}$ denote the standard Borel $\sigma-$algebra on $\mathbb{X}$, and recall that $\pi$ is the stationary distribution of $\{X_t\} \subset \mathbb{X}$. For some $p\geq 1$, we say that $L_p(\mathbb{X}, \mathcal{F}, \pi) $ is the space of (equivalence classes of) measurable functions $f$ satisfying $\int |f(x)|^p \: d\pi < \infty$. We let $L_p(\mathbb{X}, \mathcal{F}, \pi)_+$ denote the subset of these functions that are almost everywhere positive. We write $L_p(\mathbb{X})_+$ for shorthand. 

The $L_p$ norm of a function $f \in L_p(\mathbb{X})$ is given by $||f||_p = \int|f(x)|^p d\pi = \mathbb{E}_\pi |f|^p$. For each consumption process $C = \{C_t\}_{t\in\mathbb{N}}$, consider the $L_p(\mathbb{X},\pi)$ norm of a long-run mean consumption growth rate given by
\begin{equation} \label{sad}
        \lim_{n \to \infty} \,
        \left|\left| \left\{
            \mathbb{E}_{x} \left( \frac{C_n}{C_0} \right)^{1-\gamma}
        \right\} \right|\right|_p^{1/n}=
        \lim_{n \to \infty} \,
        \left( \mathbb{E}_\pi \left\{
            \mathbb{E}_{x} \left( \frac{C_n}{C_0} \right)^{1-\gamma}
        \right\}^p  \right)^{1/np}
        =
        \mathcal{M}_{C,p}^{1 - \gamma}.
\end{equation}
where $x \in \mathbb{X}$ denotes some starting state, and $\pi$ is the stationary distribution governing the process $\{X_t\}$. Proposition \ref{welldef} shows existence of this expression.
Let $\theta = \frac{1-\gamma}{1 - 1/\psi}.$ This paper's results centre around a corresponding value
$$\Lambda_p = \beta \mathcal{M}_{C,p}^{1/\theta}.$$

In the $L_1(\mathbb{X})$ setting considered in \cite{BorovickaStachurski2017}, $p = 1$ in equation \eqref{sad}. Thus, the law of iterated expectations means that
\begin{equation} \label{happy}
        \lim_{n \to \infty} \,
        \left|\left| \left\{
            \mathbb{E}_{x} \left( \frac{C_n}{C_0} \right)^{1-\gamma}
        \right\} \right|\right|_1^{1/n}
        =\lim_{n \to \infty} \,
         \left\{
            \mathbb{E}_{\pi} \left( \frac{C_n}{C_0} \right)^{1-\gamma}
        \right\}^{1/n}
        =
        \mathcal{M}_{C,1}^{1 - \gamma}.
\end{equation}
Equation \eqref{happy} is much simpler than the expression in \eqref{sad}, both in terms of numerical implementation and economic intuition. Nevertheless, considering equation \eqref{sad} will allow for a bigger set of applications and sharper solutions.

We now consider the case where $\mathbb{X}$ is a compact metric space.

\section{ Recursive Utility with Time Preference Shocks}

In this section we find existence and uniqueness conditions for recursive utilities with time preference shocks. Consistent with the approach taken in \cite{HansenScheinkman} and \cite{BorovickaStachurski2017}, we seek a normalised solution to
\begin{equation}
    G_t : = \Big( \frac{V_t}{C_t}\Big)^{1-\gamma}
\end{equation}
as it is easier to solve for $G_t$ than $V_t$.

We seek to express equation \eqref{eppy} in terms of $G_t$ to find a stationary Markov solution. By homogeneity of the aggregator $W$, we see
\begin{equation}
    \frac{V_t}{C_t} = \left\{ (1-\beta)\lambda_t + \beta \left\{ \mathcal{R}_t\Big( \frac{V_{t+1}}{C_{t+1}} \frac{C_{t+1}}{C_t} \Big)  \right\}^{1- 1/\psi} \right\}^{1/(1-1/\psi)}
\end{equation}
Using the consumption specification in assumption \ref{ass1} and definition of $\mathcal{R}_t$ yields
\begin{equation}
    \frac{V_t}{C_t} = \left\{ (1-\beta)\lambda_t + \beta  \left\{ \left[ \mathbb{E}_t \Big[ \frac{V_{t+1}}{C_{t+1}} \exp[\kappa(X_{t+1}, X_t, \epsilon_t)] \Big]^{1-\gamma} \right]^\frac{1}{1-\gamma} \right\}^{1-1/\psi} \right\}^{1/(1-1/\psi)} 
\end{equation}
Taking this expression to the power of $1-\gamma$ and rewriting in terms of $G_t$ gives the recursion
\begin{equation}
    G_t = \left\{ (1- \beta)\lambda_t + \beta \left(\mathbb{E}_t \Big[ G_{t+1} \exp[(1-\gamma) \kappa(X_t,X_{t+1}, \epsilon_{t+1})] \Big] \right)^{1/\theta}  \right\}^\theta 
\end{equation}
where
$$\theta = \frac{1-\gamma}{1 - 1/\psi}.$$
We seek a stationary Markov solution of the form $G_t = g(x)$ where $x \in \mathbb{X}$. This translates into the functional fixed point problem
\begin{equation}
    g(x) = \left\{ (1- \beta)\lambda(x) + \beta \left(\int g(x) \int \exp[(1-\gamma) \kappa(x,y, \epsilon)] \nu(d\epsilon)\: p(x,y)\:dy \right)^{1/\theta}  \right\}^\theta \label{stationarymarkov}
\end{equation}
where $\nu$ is the distribution of $\epsilon_{t+1}$, and $\lambda(x) \in C(\mathbb{X})$ is continuous.

It is convenient to express equation \ref{stationarymarkov} in terms of an operator equation. Moreover, let the preference shock  $(1-\beta)\lambda(x) = \xi(x).$ That is,
\begin{equation}\label{operator}
    A g(x) = \left\{\xi(x) + \beta \left(\int g(x) \int \exp[(1-\gamma) \kappa(x,y, \epsilon)] \nu(d\epsilon)\: p(x,y)\:dy \right)^{1/\theta}  \right\}^\theta. 
\end{equation}
In particular, the recursive utility representation has a unique solution if and only if $A$ has a fixed point. Drawing from \cite{BorovickaStachurski2017}, we may further decompose this problem into $A g(x) = \varphi(x, Kg(x))$ where
\begin{equation}\label{linear}
    Kg(x) = \int g(y) \int \exp[(1-\gamma)\kappa(x,y,\epsilon)] \nu(d\epsilon)q(x,y) dy, 
\end{equation}
and $\varphi(x,t)$ is the scalar-valued function given by
\begin{equation}
    \varphi(x,t) = \Big\{ \xi(x) + \beta t^{1/\theta} \Big\}^\theta.
\end{equation}

Note that $K$ is a linear operator in $g$. Let $\rho$ denote the spectral radius of a linear operator. This yields the following useful result.

\begin{proposition}\label{welldef}
$\Lambda_p$ is well defined and satisfies $\Lambda_p = \beta \, \rho(K)^{1/\theta}$.
\end{proposition}
\begin{proof}
See appendix A.2. This is proven as proposition \ref{p:lsr}.
\end{proof}

We take an auxiliary assumption for our main result.

\begin{assumption} \label{ass3}
The state space $\mathbb{X}$ is compact.
\end{assumption}

We can now state a central theorem of this thesis.

\begin{theorem}\label{BIG} Let $\Lambda_p = \beta \mathcal{M}_{C,p}^{1/\theta}$. Under assumptions \ref{ass1}, \ref{ass2}, \ref{ass3}, the following statements are equivalent:

a) $\Lambda_p < 1$.

b) $A$ has a fixed point in $L_p(\mathbb{X})_+$

c) There exists a $g \in L_p(\mathbb{X})_+$ such that $\{A^ng\}_{n \geq 1} $ converges to an element of $L_p(\mathbb{X})$.

d) $A$ has a unique fixed point in $L_p(\mathbb{X})$

e) $A$ has a unique fixed point, $g^* \in L_p(\mathbb{X})_+$, and $A^ng \to g^*$ as $n \to \infty$ for any $g\in L_p(\mathbb{X})_+$.
\end{theorem}

Theorem \ref{BIG} provides a necessary and sufficient condition for existence and uniqueness of recursive utilities in terms of the value $\Lambda_p$. Part (e) also establishes global stability of the solution.


\section{ Recursive Utility with Narrow Framing }

In this section, we study the existence and uniqueness of a recursive utility model where the agent has `narrow framing'. 

The \cite{Barberis} model of recursive utility with narrow framing can be written as
$$U_t = W(C_t, \mathcal{R}_t(U_{t+1}) + B_t ) $$
where $W$ is the CES aggregator and $B_t$ is a function of the state $X_t$ which captures narrow framing. 

The intuition for this is as follows. As discussed by \cite{guophd} and \cite{guo}, narrow framing is conceptually equivalent to `utility for gains and losses'. Thus, the arguments in the aggregator $W$ must change. Instead of trading off between consumption today and value tomorrow, the agent chooses between consumption today and value \textit{plus} some gain/loss tomorrow.

Once again, we use homogeneity of the aggregator to seek a normalised solution to
$$ \frac{U_t}{C_t} = \Big\{ (1 - \beta) + \beta \Big[ \mathcal{R}_t \Big( \frac{U_{t+1}}{C_{t+1}} \frac{C_{t+1}}{C_t} \Big) + \frac{B_t}{C_t} \Big]^{1-\frac{1}{\psi}} \Big\} ^\frac{1}{ 1-\frac{1}{\psi} }.  $$

We then convert this problem into a functional fixed point equation, where the left hand side is a Markovian function of the state. With regard to the framework presented earlier, this means our solution will be a fixed point to the operator defined by
$$ Bg(x) = \Big\{ (1- \beta) + \beta \Big( Kg(x) + b(x)  \Big)^\frac{1}{\theta} \Big\}^\theta  $$
where $K: L_p(\mathbb{X})_+ \to L_p(\mathbb{X})_+$ is a bounded, linear operator, $\beta \in (0,1)$, $g \in L_p(\mathbb{X})_+$ and $b \in C(\mathbb{X})$ is a strictly positive, continuous function on $\mathbb{X}$.

This mathematical framework allows us to establish sufficient conditions for the existence and uniqueness of a non-trivial solution.

\begin{proposition}\label{narrow}
Under assumptions \ref{ass1}, \ref{ass2}, \ref{ass3}, if $\Lambda_p < 1,$ then $B$ has a fixed point $ g^*\in L_p(\mathbb{X})_+$. Moreover, $B$ is globally stable in the sense that $A^ng \to g^*$ as $n \to \infty$ for any $g \in L_p(\mathbb{X})$.
\end{proposition}
For a proof of this proposition, see appendix A.3. 

\begin{figure}[H]
    \centering
    \includegraphics[scale=0.6]{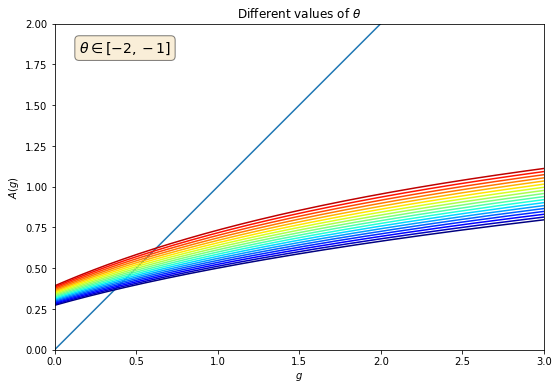}
    \caption{Fixed points of the narrow framing recursive utility operator when $\Lambda_p>1$.}
    \label{fig:1}
\end{figure}

Note that this proposition only ensures sufficiency. To see why necessity fails, consider the following counter-example. Suppose that $\mathbb{X} = x_0$ is a singleton space. Then the dimension of the problem reduces from infinity to one. In this case, $B:\mathbb{R}_+ \to \mathbb{R}_+$, and $K$ satisfies $\rho(K) = |K|$. If we choose $0<K/\beta<1$ and $\theta < 0$, then $\Lambda_p = \beta \rho(K)^{1/\theta}>1$. Nevertheless, there may still be a non-trivial fixed point which occurs due to the narrow framing term. This can be seen in figure \ref{fig:1}.

\begin{figure}[H]
    \centering
    \includegraphics[scale=0.6]{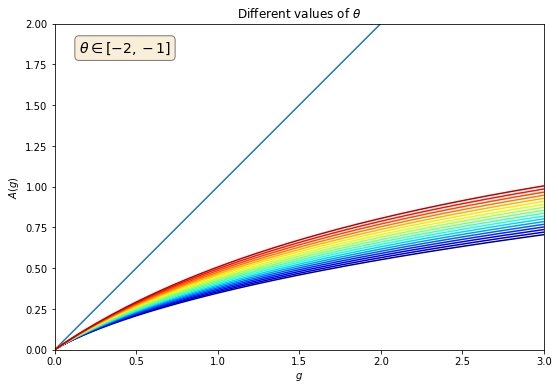}
    \caption{Without narrow framing, $ A $ has only the trivial fixed point.}
    \label{fig:2}
\end{figure}

To see why this fixed point does not occur in the standard Epstein-Zin representation, see figure \ref{fig:2}. The absence of narrow framing shifts the utility process down to avoid the fixed point.

\section{Unbounding the State Space}

In practice, models using recursive utility will be implemented numerically. When this is the case, the state space is discretized and hence finite. The state space will thus satisfy the compactness assumption in  \ref{ass3}. In theoretical models however, the state space is often unbounded. For example, \cite{bansal1} and \cite{Schorfheide} employ consumption processes which incorporate numerous unbounded shocks.

Evidently, assumption \ref{ass3} is restrictive. In the following analysis, we extend the unboundedness results of \cite{BorovickaStachurski2017}. The analysis invokes a regularity assumption known as \textit{eventual compactness} of the operator $K$.

\begin{definition}
An operator $K$ is called \textit{compact} if the closure of $K(B)$ is compact for all bounded subsets $B \subset L_p(\mathbb{X})$.
\end{definition}

\begin{definition}
An operator $K$ is called \textit{eventually compact} if there exists an $i\in \mathbb{N}$ such that $K^i$ is compact.
\end{definition}

In the $L_1(\mathbb{X})$ setting used in \cite{BorovickaStachurski2017}, operator compactness is hard to determine. The Hilbert space structure of $L_2(\mathbb{X})$ however allows more operators to satisfy compactness.\footnote{We will see why this is true in proposition \ref{L2}.} This is a primary motivation for extending the solution space from $L_1(\mathbb{X})$ to $L_p(\mathbb{X})$.

\begin{assumption}\label{eventually}
The linear operator $K$ is bounded (and hence continuous), and eventually compact. Moreover, let $\mathbb{X}$ be $\sigma-$finite.
\end{assumption}

For the next theorem, recall that $\Lambda_p = \beta \mathcal{M}_{C,p}^{1/\theta}$. We maintain the assumption that the function $\lambda(x)$ is continuous takes values on a compact set.

\begin{theorem}\label{unbounded}
Let $\mathbb{X}$ be a (possibly unbounded) metric space, and $A$ be the operator from equation \eqref{operator}. If assumptions \ref{eventually}, \ref{ass1} and \ref{ass2} hold, then the following statements are equivalent:

a) $\Lambda_p < 1$.

b) $A$ has a fixed point in $L_p(\mathbb{X})_+$.

c) There exists a $g \in L_p(\mathbb{X})_+$ such that $\{A^n g \}_{n\in \mathbb{N}}$ converges to an element of $L_p(\mathbb{X})_+$.
\end{theorem}
 
 The proof of this theorem is located in the appendix A.4. Note, we lose stability of the solution. Before moving on to applications, there is one more auxiliary result which can assist in establishing existence and uniqueness in $L_2(\mathbb{X}, \pi)$.
 
 \begin{proposition}\label{L2}
  Let the (Schwartz) kernel be
  $$ k(x,y) = \int \exp[(1-\gamma) \kappa(x,y,\epsilon)] \nu(d\epsilon) q(x,y).$$
  Let $k(x,y) \in L_2(\mathbb{X} \times \mathbb{X}, \pi \times \pi)$. Then the linear operator $K$, defined in equation \eqref{linear}, is compact in $L_2(\mathbb{X}, \pi)$.\footnote{This compactness result is unique to $L_2$, and does not hold in general for other $L_p$ spaces. This is because $L_2$ is a Hilbert space, and the proof of compactness relies on an approximation argument relying on an orthonormal basis expansion.}
 \end{proposition}
 \begin{proof}
 Substituting the expression for $k(x,y)$ into $Kg(x)$ gives
\begin{equation} \label{hilbertschmidtoperator}
Kg(x) = \int g(y) k(x,y) dy. 
\end{equation}
If $k(x,y) \in L_2(\mathbb{X} \times \mathbb{X}, \pi \times \pi)$, then $K$ is a Hilbert-Schmidt integral operator in $L_2(\mathbb{X},\pi)$. Hilbert-Schmidt operators are compact: see page 198 of \cite{stein} for a full proof.
 \end{proof}

\clearpage

\chapter{ Applications and Simulations }
\textit{``I don't give a damn about the applications - show me the equations.'' - John Stachurski}


\section{Bansal-Yaron with Constant and Stochastic Volatility}
We now show existence and uniqueness of the recursive utility specification seen in section I.A of \cite{bansal1}. We also consider a truncated `stochastic volatility' model from section I.B. All results are new contributions which consider unbounded cases.

Despite the broad success of \cite{bansal1} in explaining asset pricing puzzles, unique existence of a solution has not yet been resolved. The closest result is \cite{pohl}, which only proves the existence of a solution. \cite{pohl} impose a stringent bounded condition on the parameters. By contrast, we establish uniqueness \textit{as well as} existence. Further, the conditions we impose are less strict in that they are not only sufficient, but necessary, for a unique solution.

\cite{bansal1} represent preferences with the standard Epstein-Zin recursion
$$ V_t = \left[ (1-\beta)C_t^{1-1/\psi} +\beta \{ \mathcal{R}_t(V_{t+1}) \}^{1-1/\psi}  \right]^{1/(1-1/\psi)} $$
as seen in earlier sections.\footnote{Note that this is the degenerate form of the the \cite{Schorfheide} preference representation. Set $\lambda(x) = 1$.} In this model, consumption grows according to
\begin{equation}
    \ln(C_{t+1}/C_t) = \mu_c + z_t + \sigma \eta_{c,t+1}
\end{equation}
 where $\{ \eta_{c,t+1} \}$ are IID standard normal. 
 
 In section I.A of their paper, \cite{bansal1} capture stochastic growth via the autoregressive process
 \begin{equation}
     z_{t+1} = \rho z_t + \sigma \eta_{z,t+1}
 \end{equation}
 where the innovation process $\{ \eta_{z,t+1} \}$ is IID standard normal. We represent the state process by $X_t = x$. The central linear valuation operator can thus be written as:
 \begin{align}
    Kg(x) &= \int g(y) \int \exp\Big[ (1-\gamma) \kappa(x,y,\epsilon)\Big] \nu(d\epsilon) q(x,y) dy \\
    & = \int g(y) \int \exp\Big[(1-\gamma)(\mu_c+ x_1 + \sigma \epsilon)\Big] \nu(d\epsilon) q(x,y) dy . \label{bansalA_operator}
\end{align}
Notably, the state space is given by $\mathbb{X} = \mathbb{R}$.

\begin{proposition}\label{HilbertSchmidtA}
 The operator $K$, defined in equation \eqref{bansalA_operator}, is compact in $L_2(
 \mathbb{X}, \mathcal{F}, \pi).$
\end{proposition}
\begin{proof}
Observe that
$$k(x,y) = \int \exp\Big[ (1- \gamma) (\mu_c + x_1 + \sigma \epsilon) \Big] \nu(d\epsilon) q(x,y). $$

First, note
\begin{align*}
    k(x,y) & = \int \exp [(1-\gamma) (\mu_c + x_1 + \sigma \epsilon)] \nu(d\epsilon) q(x,y) \\
    & = \exp\Big[(1-\gamma)(\mu_c + x) + (1-\gamma) \sigma^2 / 2 \Big] q(x,y) \\
    & = \frac{1}{\sqrt{2\pi} \sigma} \exp\Big[(1-\gamma)(\mu_c + x) + (1-\gamma)^2 \sigma^2 / 2 \Big] \cdot \exp\Big[{-(y-\rho x)^2 / 2 \sigma^2}\Big] 
\end{align*}
Further, the stationary distribution of $\{z_t\}$ is given by $\pi(x) = \frac{\sqrt{(1-\rho^2)}}{\sqrt{2\pi} \sigma} \exp[ -x^2(1-\rho^2)/2\sigma^2 ]$. Thus, using Fubini's theorem over the product measure $\pi^2 = (\pi \times \pi)(x \times  y)$, we obtain
\begin{align*}
    \int |k(x,y)|^2 d\pi^2 & = \frac{1}{2 \pi \sigma^2}\int e^{ 2(1-\gamma)(\mu_c + x) +(1-\gamma)^2 \sigma^2 - (y-\rho x)^2 }    \: d\pi^2\\
    & = \frac{e^{2(1-\gamma)^2\sigma^2}}{2\pi \sigma^2} \int e^{ 2(1-\gamma)(\mu_c + x)}  \int e^{ -(y-\rho x)^2 / \sigma^2 }   d\pi(y) d\pi(x)\\
    & \leq \frac{e^{2(1-\gamma)^2\sigma^2}}{\sqrt{2\pi} \sigma} \int e^{ 2(1-\gamma)(\mu_c + x)} d\pi(x)\\
    &= \frac{e^{2(1-\gamma)^2\sigma^2}\sqrt{(1-\rho^2)} }{2\pi \sigma^2} \int e^{2(1-\gamma)(\mu_c +x)} e^{-x^2(1-\rho^2)/2\sigma^2} \: dx \\
    & = \frac{e^{2(1-\gamma)^2\sigma^2 +2(1-\gamma)\mu_c}\sqrt{(1-\rho^2)} }{2\pi \sigma^2} \int e^{2(1-\gamma)x}e^{-x^2(1-\rho^2)/2\sigma^2} \: dx \\
    & = \frac{e^{2(1-\gamma)^2\sigma^2 +2(1-\gamma)\mu_c } \sqrt{1-\rho^2}  }{\sqrt{2\pi} \sigma^2 } \cdot e^{4(1-\gamma)^2\frac{\sigma^2}{2(1-\rho^2)}} \\
    &<\infty.
\end{align*}
That is, the Schwartz kernel $k(x,y)$ is bounded in $L_2(\mathbb{R})$. Applying proposition \ref{L2} shows that $K$ is compact in $L_2(\mathbb{X},\pi)$.
\end{proof}

Proposition \ref{HilbertSchmidtA} and theorem \ref{unbounded} show that the constant volatility model of \cite{bansal1} has a unique solution if and only if $\Lambda_2 <1$. Before moving on to see example parameterisations, we now consider the stochastic volatility case.

 In section I.B of \cite{bansal1}, the authors employ the dual laws of motion
\begin{equation}\label{mean}
    z_{t+1} = \rho z_t + \varphi_e \sigma_t \eta_{z,t+1} 
\end{equation}
\begin{equation}\label{variance}
    \bar{\sigma}^2_{t+1} = \nu \bar{\sigma}_t^2 + d + \varphi_\sigma \eta_{\sigma,t+1}
\end{equation}  
where the innovation process $ \left\{ \eta_{i,t} \right\} $ is IID standard normal for $i \in \{z, \sigma \}$. The state vector $X_t$ can be represented as $X_t = (z_t,\sigma_t^2)$ with $x = (x_1,x_2)$. In order to  make $\sigma_t$ well-defined in $\mathbb{R}$, we define it by
$$ \sigma_t = \bar{\sigma}_t\mathbb{1}_{\bar{\sigma}_t^2 \geq 0} - i \: \bar{\sigma }_t \mathbb{1}_{ \bar{ \sigma }_t^2 < 0}.   $$
where $i \in \mathbb{C}$ is the imaginary unit.

The operator $K$ can thus be written as
\begin{align}
    Kg(x) &= \int g(y) \int \exp\Big[ (1-\gamma) \kappa(x,y,\epsilon)\Big] \nu(d\epsilon) q(x,y) dy \\
    & = \int g(y) \int \exp\Big[(1-\gamma)(\mu_c+ x_1 + \sqrt{x_2} \epsilon)\Big] \nu(d\epsilon) q(x,y) dy\\ \label{bansaloperator}
    & = \int g(y) k(x,y) dy.
\end{align}
 Here, the state space is given by $\mathbb{X} = \mathbb{R}^2$. 
 
 The stationary distribution of $\{z_t\}_{t\in \mathbb{N}}$ is inconvenient; it is the product of two independent Gaussian random variables. The corresponding probability density function is too heavy tailed to establish existence. It decays asymptotically on the order of $f(X = x) \sim x^{-1/2}e^{-hx}$.\footnote{ This is demonstrated in \cite{bernoulli}. } This means that bounding the kernel, $k(x,y)$, is generally intractable on an unbounded state-space. 
 
 
 To overcome the heavy tail problem, we truncate the shocks to stochastic volatility but leave other shocks unbounded. That is, assume $\{\eta_{\sigma,t+1}\}$ are bounded and that $\{ \sigma_t \}$ thus converges to some \textit{bounded} distribution.\footnote{As opposed to unbounded Gaussian as per the original model.} The Bansal-Yaron dynamical system thus evolves according to
 \begin{equation}\label{approxmean}
    z_{t+1} = \rho z_t + \varphi_e (\sigma_t + \epsilon) \eta_{z,t+1} 
\end{equation}
\begin{equation}\label{approxvariance}
    \bar{\sigma}^2_{t+1} = \nu \bar{\sigma}_t^2 + d + \varphi_\sigma \eta_{\sigma,t+1}
\end{equation}  
for some $\epsilon > 0$. The inclusion of $\epsilon > 0$ is for mathematical convenience; it stops the innovation terms in \eqref{approxmean} from degenerating.

\begin{proposition}\label{approxthm}
Let be $\{ \eta_{\sigma,t+1}\}$ be uniformly bounded with absolutely continuous density, such that $x_2 \in [0,M] $ for some $M>0$. Then the linear operator $K$ is compact in the stochastic volatility Bansal-Yaron model.
\end{proposition}
\begin{proof}
This proof builds upon the argument seen in the proof of proposition \ref{HilbertSchmidtA}. See appendix A.5 for the full proof.

\end{proof}

Proposition \ref{HilbertSchmidtA} implies that we may apply theorem \ref{unbounded} to the section I.A \cite{bansal1} model. For constant volatility, this shows that the utility representation is well defined if and only if the parameters satisfy $\Lambda_2 < 1$. For stochastic volatility, proposition \ref{approxthm} shows that the model with truncated `uncertainty' is also well defined if and only if $\Lambda_2 < 1$.

The \cite{bansal1} parameterisation satisfies this condition for a unique solution (see table \ref{table1}). We demonstrate this numerically. We treat the stochastic volatility case since it is the more groundbreaking model.


\begin{table}[h]
    \centering
    \begin{tabular}{ c c c c c c c c}
    $\mu_c$ & $\rho$  & $\bar \sigma$ & $\varphi_e$ & $\nu$ & $\varphi_\sigma$ & $\psi$ & $\gamma$ \\
    \hline
    0.0015 & 0.979 & 0.0078 & 0.044 & 0.987 & 2.3 $\cdot$ 10$^{-6}$ & 1.5 & 10
    \end{tabular}
    \caption{ \cite{bansal1} parameter values for $\beta = 0.998$. }
    \label{table1}
\end{table}

Here, the numerical implementation required to find $\Lambda_2$ is different to that required in \cite{BorovickaStachurski2017}. In the latter paper, the authors need only test $\Lambda_1$ because the state space is compact. In \cite{BorovickaStachurski2017}, the authors show that $\Lambda_1 \approx 0.998$.

\begin{figure}[H]
    \centering
    \includegraphics[scale=0.6]{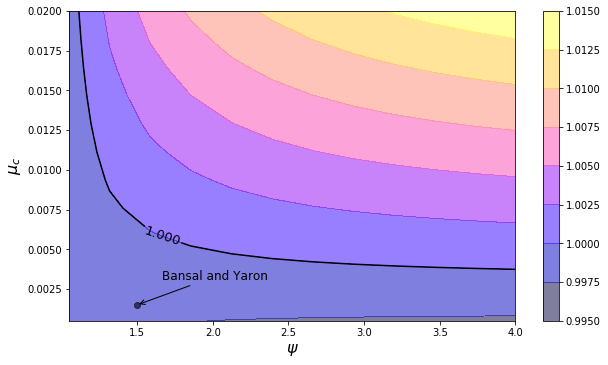}
    \caption{Stability Map for the Bansal-Yaron Model}
    \label{fig:3}
\end{figure}

In our approach, we implement a Monte Carlo method to first estimate
\begin{align}
   h(x) &= \mathbb{E}_x \Big( \frac{C_n}{C_0} \Big)^{1-\gamma} \\
   &\approx \left[ \frac{1}{m} \sum^m_{j=1} \Big( \frac{C_n^j}{C_0^j} \Big)^{1-\gamma}  \right]
\end{align}
for 1000 draws of $x\sim U(0, 100)$. We then approximate
$$ \Big(\int h(x) \: d\pi(x)\Big)^{1/2n}. $$
This process can also be implemented by choosing $x$ from the stationary distribution, $\pi$, and then estimating a new integral
$$ \Big(\int h(x) \: d\pi(x)\Big)^{1/2n} = \Big(\int h_\pi (x) \: dx\Big)^{1/2n} \approx \left( \frac{1}{m} \sum^m_{i=1} h(x_i) \right)^{1/2n}. $$
Choosing $m = n = 1000$ yields $\Lambda_2 \approx 0.998$. This suggests that even on an unbounded state space, the original Bansal-Yaron model is well defined. This can also be seen in figure \ref{fig:3}, where the Bansal-Yaron parameterisation is safely within the stability zone for values of $\psi$ and $\mu_c$.\footnote{ This result can be backed up by a heuristic mathematical argument. Note, that the process $\{ z_t \}$ has a stationary distribution in $L_2(\mathbb{X})$. Thus, the growth rate of $C_n$ for large $n$ taken from this stationary distribution is almost surely bounded. Averaging the draws of $C_n$ and taking to the power of $1/n$ then makes this value arbitrarily close to $1$ when the exponent $1/n$ `dominates' the growth rate. In this case, multiplying by the discount $0.998$ means that the final integrated value will be almost surely arbitrarily close to $0.998$.}

\section{ Mehra-Prescott [1985] and Epstein-Zin [1990]}

The seminal models of \cite{MehraPrescott} and \cite{Epstein1990} employ a consumption specification with permanent innovations of the form
$$ \ln\Big(\frac{C_{t+1}}{C_0}\Big) = t \ln(1+g) + \xi_{t+1}$$
where $ \xi_{t+1} = (1-a) + a\xi_t + u_{t+1}$, for $a \in (0,1]$ and $u_{t+1}$ IID standard normal. The state process here is determined by setting $\{X_t\} = \{ \xi_t \}.$

In terms of the state process, $\{X_t\}$, we can write the consumption growth rate as
\begin{align}
    \ln\Big( \frac{C_{t+1}}{C_t}\Big) = \ln(1+g) + (1-a) +(a-1)X_t  + u_{t+1} \label{mehraspec}
\end{align}

\begin{proposition}\label{applicationshit}
 The linear operator $K$, corresponding to equation \eqref{mehraspec}, is compact.
\end{proposition}
\begin{proof}
Note that equation \eqref{mehraspec} is of a form similar to Bansal and Yaron with constant volatility. Thus, the proof of compactness follows a similar logic. 
\end{proof}
Applying proposition \ref{applicationshit}, we can see that \cite{MehraPrescott} and \cite{Epstein1990} have an $L_2$ solution precisely when $\Lambda_2 < 1$.

\section{Schorfheide, Song and Yaron [2018]}

Consumption in the \cite{Schorfheide} model is determined by the following state dynamics
\begin{align}
    \ln (C_{t+1} /  C_t) &
    = \mu_c + z_t + \sigma_{c, t} \, \eta_{c, t+1},
    \label{al:ssyc1}
    \\
    z_{t+1} = \rho \, z_t
        & + \sqrt{1 - \rho^2} \, \sigma_{z, t} \, \eta_{z, t+1},
    \label{al:ssyc2}
    \\
    \sigma_{i, t} = \phi_i \, \bar{\sigma} \exp(h_{i, t})
    & \quad \text{with} \quad
    h_{i, t+1} = \rho_{i} h_i + \sigma_{h_i} \eta_{h_i, t+1},
    \quad i \in \{c, z\}.
    \label{al:ssyc3}
\end{align}

Here, $\{\eta_{i, t}\}$ and $\{\eta_{h_i, t}\}$ are {\sc iid} and
standard normal for $i \in \{c, z\}$.  The associated state vector can be represented as $X_t = (h_{c, t}, h_{z,t}, z_t)$, where $x = (x_1, x_2, x_3)$. As discussed earlier, the Schorfheide-Song-Yaron model ranks consumption streams according to
\begin{equation}
    V_t = \Big[ (1 - \beta) \lambda_t C_t^{1-1/\psi} + \beta \{ \mathcal{R}_t(V_{t+1}) \}^{1-1/\psi} \Big]^{1/(1-1/\psi)}
\end{equation}

There are two factors which complicate existence in this model. First, \cite{Schorfheide} augment the Epstein-Zin model of recursive utility with the time preference shock, $\lambda_t$. This issue was solved by theorem \ref{BIG}. Second, the model of \cite{Schorfheide} involves an unbounded stochastic volatility process. Consequently, there are very few direct results regarding existence and uniqueness of this model. 

Theorem \ref{BIG} shows that on a compact state space, the \cite{Schorfheide} model still has a unique, globally attracting solution in $L_p$. We now consider a version of the unbounded case in $L_2$. In this model there are a number of unbounded shocks at play. As we now demonstrate, we only need to impose assumptions on the stochastic volatility to get uniqueness and existence. 

\begin{proposition}\label{HilbertSchmidt2}
 Let the shock processes $\{\eta_{z,t+1}\}$, $\{\eta_{c,t+1} \} $ be uniformly bounded with absolutely continuous density such that $x_1, x_2 \in [-M, M]$ for some $M> 0$. Then the linear operator $K$ is compact for the Schorfheide-Song-Yaron specification in $L_2$.
\end{proposition}
\begin{proof}
See appendix A.5
\end{proof}

Using proposition \ref{HilbertSchmidt2}, we may apply theorem \ref{unbounded} to the Schorfheide-Song-Yaron model. Consequently, a unique solution exists if and only if the model parameters satisfy $\Lambda_2 < 1$. The next two figures illustrate this.

\begin{figure}[H]
    \centering
    \includegraphics[scale=0.6]{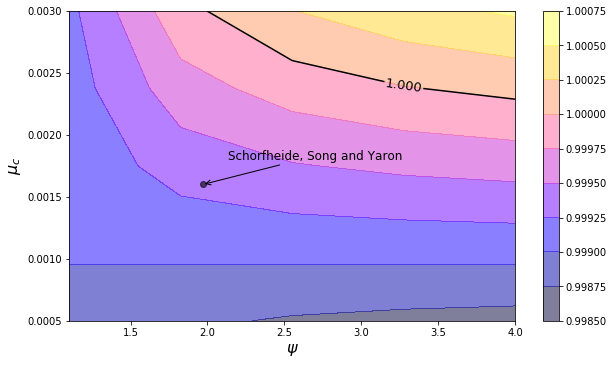}
    \caption{Values of $\Lambda_2$}
    \label{ssy:1}
\end{figure}

\begin{figure}[H]
    \centering
    \includegraphics[scale=0.6]{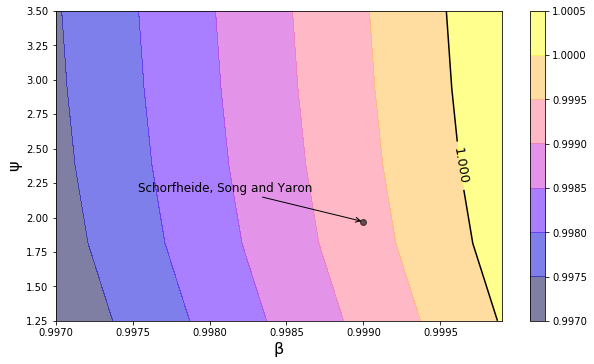}
    \caption{Values of $\Lambda_2$}
    \label{ssy:1}
\end{figure}

\clearpage

\chapter{Discussion}
\textit{ ``Why would you microfound macro when there are real problems to work on?'' - Damien Eldridge }

 \cite{BorovickaStachurski2017} express recursive utility as the composition of an infinite dimensional linear operator and a real valued function. This allows the authors to combine Perron-Frobenius theory with monotone concave operator theory. In doing so, they obtain sharp results regarding the existence and uniqueness of Epstein-Zin utilities. This thesis extends the result of  \cite{BorovickaStachurski2017}. Our contribution is twofold.
 
First, this paper establishes existence and uniqueness conditions for recursive utilities with (i) time preference shocks and (ii) narrow framing. These results are particularly useful in modern asset pricing, where theorists are increasingly relying upon more complex recursive utility representations. 
 
 Second, by altering the solution space from $L_1$ to $L_p$, we add greater flexibility to the existence and uniqueness conditions. More specifically, if a solution can't be established in one space, our results allow the practitioner to test another space. This is demonstrated in chapter 4, where we use $L_2$ to tackle unbounded \cite{bansal1} and \cite{Schorfheide} environments.

There are some shortcomings to this paper. The extension from $L_1$ to $L_p$ makes numerical implementation more costly. Moreover, all unbounded results rely on the eventual compactness of the operator $K$. But this assumption is not satisfied in heavy tailed models. As such, future research could try to relax this condition. This is non-trivial however.

\appendix
\chapter{Appendix}
\textit{``If you drop out of theory, there is always applied micro.'' - Sander Heinsalu }

\section{General Mathematical Results}

In this section, we set out the general fixed point and spectral radius results required for this paper. Where results are taken from \cite{BorovickaStachurski2017}, proofs are generalised from $L_1(\mathbb{X})$ to either general Banach spaces or $L_p(\mathbb{X})$ spaces.

Let $E$ be a Banach space over $\mathbb{R}$, and denote the zero element by $\Vec{0}$. 

\begin{definition}
A nonempty, closed, convex set $P \subset E$ is called a \textit{cone} if

1) $x \in P$, $\lambda \geq 0$ implies that $ \lambda x \in P$

2) $x \in P$ and $-x \in P$ implies that $ x = \vec{0}.$

The cone is called \textit{normal} if there exists a constant $\delta > 0 $ such that $||x + y|| \geq \delta $ for all $x, y \in P$ satisfying $||x|| = ||y|| = 1$.
\end{definition}

A cone $P \subset E$ induces a partial ordering $\leq$, by defining
\begin{equation}\label{ordering}
u \leq v \iff v - u \in P.
\end{equation}
We can take a strict ordering by also requiring that $u - v \not\in P$. A Banach space with this cone-order structure may then be called a \textit{partially ordered }Banach space. The cone generating the partial ordering is called a \textit{positive} cone. Further, let $D \subset E$ and define the operator $A:D \to E$. We say that $A$ is an \textit{increasing} (or isotone) operator if for all $x_1,x_2 \in D$, where $x_1 \leq x_2$, we have $Ax_1 \leq Ax_2$.

The next theorem is needed to establish basic fixed points results.

\begin{theorem}[\cite{Du1990} or \cite{zhang2013} Theorem 2.1.2]\label{du2}
Suppose that the cone $P$ is normal, $u_0, v_0 \in E$ and $u_0 < v_0$. Let $A : [u_0,v_0] \to E$ denote an increasing operator. If one of the following holds:

i)  $A$ is a concave operator, $Au_0 \geq u_0 + \epsilon (v_0 - u_0)$, $Av_0 \leq v_0$ where $\epsilon \in (0,1)$ is a constant;

ii) $A$ is a convex operator, $Au_0 \geq u_0$, $Av_0 \leq v_0 - \epsilon (v_0 - u_0)$ where $\epsilon \in (0,1)$ is a constant, 

then $A$ has a unique fixed point $x^*$ in $[u_0,v_0].$ Further, for any $x_0 \in [u_u,v_0],$ the iterative sequence $\{x_n\} $ given by $x_n = Ax_{n-1}$ satisfies
\begin{align*}
& ||x_n - x^*|| \leq M (1-\epsilon)^n \:\:\:\:\:\:\:\:\:\:\: (n = 1,2,...)
\end{align*}
for some $M\in \mathbb{R}_{+}$ independent of $x_0$.
\end{theorem}
\begin{proof}
See \cite{zhang2013} theorem 2.1.2.
\end{proof}

Let $\mathbb{X}$ be a compact metric space, and $\mu$ be a measure. The set of Borel measurable functions $g:\mathbb{X} \to \mathbb{R}$, such that $||g||_p = \int |g|^p d\mu < \infty $, is canonically denoted by $L_p(\mathbb{X}, \mathcal{F},\mu)$. We write $L_p(\mathbb{X})$ when it is clear which measure is being used. The dual space of $L_p(\mathbb{X})$ is a Banach space identified with $L_q(\mathbb{X})$ when $p, q$ satisfy $\frac{1}{p} + \frac{1}{q} = 1$. Note that for $g \in L_\infty(\mathbb{X})$, we naturally define $||g||_\infty = \sup\{ |g(x)| : \: x \in \mathbb{X} \} $. For $g, h \in L_p(\mathbb{X})$, we define $g \leq h$ to mean that $g(x) \leq h(x)$ for $\mu$-almost every $x \in \mathbb{X}$. Similarly, $g \ll h$ means that $g(x) < h(x)$ $\mu$-almost everywhere. These order-relations are induced by the positive cone $L_p(\mathbb{X}, \mathcal{F}, \pi)_+$ as per equation \eqref{ordering}.

Let $K:L_p(\mathbb{X}) \to L_p(\mathbb{X})$ be a linear operator. Define the \textit{operator norm} and \textit{spectral radius} respectively by $||K|| = \sup\{ ||Kg||_p : g \in L_p, ||g||_p = 1 \}$ and $\rho(K) = \sup\{ \lambda \: : \lambda \text{ satisfies } \: Kg = \lambda g \: \text{ for some } g \in L_p(\mathbb{X}) \} $. Gelfand's formula states that $\rho(K) = \lim_{n \to \infty} ||K^n||_1^\frac{1}{n}$. We say that $K$ is \textit{positive} if $Kg \geq 0$ whenever $g \geq 0$. It is \textit{bounded} if $||K||$ is finite, and \textit{compact} if the closure of $K(B)$ is compact for all bounded subsets $B \subset L_p(\mathbb{X})$.

By treating elements $f, g \in L_p(\mathbb{X})$ as points in a partially ordered Banach space, we get the usual notions of convexity and concavity for an operator $A: L_p(\mathbb{X}) \to L_p(\mathbb{X})$.

The next proposition generalises an argument found in \cite{usydnotes}.

\begin{proposition}[A general Neumann series result.]\label{neumann}
Let $K$ be a linear operator, $\lambda \in \mathbb{R}$ and $h,f \in L_p(\mathbb{X})$. If $\rho(K) <\lambda $ then the operator equation $\lambda h = Kh + f$ has a unique solution given by $h = (\lambda I - K)^{-1} f$. In particular, this value exists if and only if the geometric series expression $h = \sum^\infty_{n=0} \lambda^{-(n+1)}K^n f $ converges in norm.
\end{proposition}
\begin{proof}
 Write the equation $\lambda h = K h + f$ as $\lambda I h = K h + f$. Rearranging gives $h = (\lambda I - K)^{-1} f$. We wish to see when this expression is well defined. As such, note
 \begin{align*}
     (\lambda I - K)^{-1} & = \frac{1}{\lambda} \Big( I - \frac{K}{\lambda}\Big)^{-1} \\
     & =  \frac{1}{\lambda} \Big( I +  \frac{K}{\lambda} +  \Big( \frac{K}{\lambda} \Big)^2 + ... \Big) \:\:\:\:\:\:\:\:\: \text{ (see \cite{usydnotes} for details)}\\
     & = \sum_{n = 0} ^\infty \lambda^{-(n+1)} K^n.  
 \end{align*}
 Thus, $h = \sum^\infty_{n=0} \lambda^{-(n+1)} K^n f$. Gelfand's formula shows that this expression is well defined if $\rho(K) < \lambda$.
\end{proof}

The next lemma is stated without proof in \cite{krasnosel2012approximate}.

\begin{lemma} \label{krasnos}
Suppose $K$ is a positive linear operator with $Kh \leq \delta h$ for some $\delta >0$, and $h \in P$ where $P$ is a normal cone. If $K$ is compact and $h$ is a quasi-interior element of $P$, we have $\rho(K) \leq \delta$.
\end{lemma}
\begin{proof}
 Observe that if $K$ is a compact linear operator, then so too is the adjoint $K^*$. Hence, $\rho(K) = \rho(K^*)$ must be attained by some $f \in E$ and $f^* \in E^*$ respectively.  Thus $A^*( f^*) = \rho(A) f^*$ for some $f^* \in E^*$. By the assumption that $h$ is quasi-interior, we have $f^*(h) > 0$ for all $f^* \in P$ and so $\rho(K) = \frac{K^*f^*(h)}{ f^*(h) }$ is well defined. Consequently,
 \begin{align*}
     \rho(A) &= \frac{K^* (f^*(h))}{f^*(h)} \\
     & = \frac{f^*(K h)}{f^*(h)} \\
     & \leq \frac{ f^*(\delta h )}{ f^*(h)} \\
     & = \delta.
 \end{align*}
 This completes the proof.
\end{proof}

\begin{definition}
Let $E$ be a Banach space, $K:E \to E$ a bounded linear operator and $h \in E$. We define
$$ \rho(K,h) = \limsup_{n \to \infty} ||K^nh||^\frac{1}{n} $$
as the \textit{local} spectral radius of $K$ at $h$.
\end{definition}

The next three lemmas are local spectral radius results.

\begin{lemma} \label{basiclocal}
Let $h \in E$. If $K$ is a bounded linear operator, then
$$ 0 \leq \rho(K,h) \leq \rho(K). $$
\end{lemma}
\begin{proof}
The inequality $0 \leq \rho(K,h)$ is immediate. Moreover, by Cauchy-Schwarz
\begin{align*}
\rho(K,h) & = \limsup_{n\to \infty} ||K^nh||\\
& \leq \limsup_{n\to \infty} ||K^n||^\frac{1}{n} ||h||^\frac{1}{n} \:\:\:\:\:\:\: \text{ (by Cauchy-Schwarz)}\\
& = \limsup_{n\to \infty} ||K^n||^\frac{1}{n} \cdot  \lim_{n \to \infty} ||h||^\frac{1}{n} \\
& = \limsup_{n\to \infty} ||K^n||^\frac{1}{n}\\
& = \rho(K).
\end{align*}
This completes the lemma.
\end{proof}

\begin{lemma}\label{Danevs1} The local spectral radius satisfies the following three properties.

(1) $\rho(K,K^n h) = \rho(K,h)$ for all $h \in E$ and $m \in \mathbb{N}$ \label{danevs1.1}

(2) $\rho(aK,bh) = |a| \rho(K,h)$ for all $h \in L_1(\mathbb{X}$, $b \neq 0$ and $a \in \mathbb{R}$ \label{danevs1.2}

(3) $\rho(K,f+h) \leq \max \Big\{ \rho(K,h), \: \rho(K,f) \Big\}$ \label{danevs1.3}

\end{lemma}
\begin{proof}
Regarding (1),  this follows immediately as $\limsup_{n \to \infty} ||K^{n+m}h||^\frac{1}{n} =\limsup_{n \to \infty} ||K^{n}h||^\frac{1}{n} $

For (2), simply observe that
\begin{align*}
    \rho(aK, bh) & = \limsup_{n \to \infty} ||(aK^n )bh||^\frac{1}{n} \\
    & \leq \limsup_{n \to \infty} |a| \: |b|^\frac{1}{n} ||K^nh||^\frac{1}{n} \\
    & = |a| \rho(K,h)
\end{align*}

For (3), take an arbitrary $c \in \mathbb{R}_+$. By the definition of local spectral radius, we may choose an $m_c \in \mathbb{N}$ such that for all $n \geq m_c$, 
$$||K^n h || \leq \Big( \rho(K,h) + c \Big)^n \text{ and } ||K^n f || \leq \Big( \rho(K,f) + c\Big)^n.$$
From this, observe that
\begin{align*}
\Big|\Big| K^n \Big( \frac{h+f}{2}\Big) \Big| \Big| & \leq \frac{1}{2} \Big( \rho(K,h) + c \Big)^n + \frac{1}{2}\Big( \rho(K,f) + c \Big)^n\\
& \leq \Big( \max \Big\{ \rho(K,h), \: \rho(K,f) \Big\} + c \Big)^n.
\end{align*}
Taking $m_c \to \infty$, we can take $c \to 0$. This completes the proof. 
\end{proof}

The next lemma denotes the linear span of a set $N$ by $\text{span}\{N\}$.

\begin{lemma}[\cite{danevs1987local}]\label{Danevs2}
Let $N \subset E$. Then
\begin{align*} \sup \Big\{  \rho(K,h) : h \in N \Big\} =  \sup \Big\{  \rho(K,h) : h \in \operatorname{span}(N) \Big\} \end{align*}
\end{lemma}
\begin{proof}
Since $N \subset \text{span}(N)$ we get for free that $\sup \Big\{ \rho(K,h) \: : \: h \in N\Big\} \leq \sup \Big\{\rho(K,h) \: : \: h \in \text{span}(N)  \Big\}$. By fixing an $h \in \text{span}(N)$, we may write $ h = \sum^n_{i=1} t_i h_i$, $h_i \in N$ for $t_i \in \mathbb{R}$. By \eqref{danevs1.2} property (2) and \eqref{danevs1.3} property (3), we see
\begin{align*}
\rho(K,h) & \leq \max_i \Big\{ \rho(K, t_i h_i) \Big\}\\
& \leq \max_i \Big\{ \rho(K,h_i) \Big\} \\
& \leq \sup_{f \in N} \Big\{ \rho(K,f) \Big\}
\end{align*}
This completes the proof.
\end{proof}

This cohort of lemmas gives rise to the final corollary.

\begin{corollary}[\cite{danevs1987local}]\label{Danevs3}
Let $N \subset E$. Then
$$ \sup \Big\{  \rho(K,h) \: : \: h \in N \Big\} = \sup \Big\{  \rho(K,h) \: : \: h \in  M \Big\} $$
where $M = \operatorname{span} \{ K^mh \: : \: h \in N, m \geq 0 \}$ 
\end{corollary}
\begin{proof}
By  \ref{Danevs1} (1), we have $\rho(K, K^nh) = \rho(K,h)$ and so $ \sup\{\rho(K,h) \: : \: h \in N \} = \sup\{ \rho(K,K^nh) \: : \: h \in N \} $. Then by \ref{Danevs2}, we also have $\sup\{ \rho(K,K^nh) \: : \: h \in N \} = \sup\{ \rho(K,K^nh) \: : \: h \in \text{span}_{h\in N}(K^n h) \} = \sup \{ \rho(K,h) \: h \in \text{span}_{h\in N} (K^nh)\}$
\end{proof}

An $L_1$ version of the following local spectral radius theorem is found in \cite{BorovickaStachurski2017}, theorem A.1. 

\begin{theorem}[Zabreiko--Krasnosel'skii--Stetsenko--Zima, \cite{BorovickaStachurski2017}]
    \label{t:lsr}
    Suppose $h \in L_p(\mathbb{X})$. Let $K$ be a positive compact
    linear operator.  If $h \gg 0$, then
    \begin{equation}
        \label{eq:lsr}
        \rho(K,h) = \lim_{n \to \infty} \| K^n h \|^{1/n} = \rho(K).
    \end{equation}
\end{theorem}

\begin{proof}[Proof]
    In  lemma \ref{basiclocal}, we established that $\rho(K, h) \leq \rho(K)$. Thus, it suffices to show that
    $\rho(K, h) \geq \rho(K)$.  Let $\lambda$ be a constant
    satisfying $\lambda > \rho(K, h)$ and let
    \begin{equation}
        \label{eq:xl}
        h_\lambda := \sum_{n=0}^\infty \frac{K^n h}{\lambda^{n+1}}.
    \end{equation}
    The point $h_\lambda$ is a well-defined element of $L_p(\mathbb{X})_+$ by
        $\limsup_{n \to \infty} \| K^n h \|^{1/n} < \lambda$
    and the Cauchy Root Test for convergence.  It is also positive $\mu$-almost
    everywhere since
    the sum in expression \eqref{eq:xl} includes $h \gg 0$ and because $K$ is a positive
    operator.  Applying \ref{neumann}, the point $h_\lambda$
    also has the representation $h_\lambda = (\lambda I - K)^{-1} h$,
    from which we obtain $\lambda h_\lambda - K h_\lambda = h$.  Since
    $h \in L_p(\mathbb{X})_+$ and $h > 0$, this implies that
         $K h_\lambda \leq \lambda h_\lambda$.
    Accordingly, by the compactness of $K$,
    quasi-interiority of $h_\lambda$ and lemma \ref{krasnos}, we must have $\rho(K) \leq \lambda$.
    Since this inequality was established for an arbitrary $\lambda$
    satisfying $\lambda > \rho(K, h)$, we conclude that $\rho(h, K) \geq \rho(K)$.
    Hence $\rho(K, h) = \rho(K)$.
    Finally, since $K$ is compact, corollary \ref{Danevs3}
    implies that $\rho(K, h) = \lim_{n \to \infty} \| K^n h \|^{1/n}$, so
    equation \eqref{eq:lsr} holds.
\end{proof}

The next result is an extension of theorem~\ref{t:lsr}. The $L_1$ version is due to \cite{BorovickaStachurski2017}.

\begin{theorem}
    \label{t:lsr2}
    Suppose $h \in L_p(\mathbb{X})$ and let $K$ be a
    linear operator on $L_p(\mathbb{X})$.
        If $K^i$ is compact for some $i \in \mathbb{N}$ and
        $K f \gg 0$ whenever $f \in L_p(\mathbb{X})_+$, then
    \begin{equation}
        \label{eq:lsr2}
         \rho(K) = \lim_{n \to \infty}
        \Big\{ \int |K^n h |^p\: d \mu \Big\}^{\frac{1}{pn}}.
    \end{equation}
    for all $h \gg 0$.
\end{theorem}

\begin{proof}
    Fix $h \in L_p(\mathbb{X})$ with $h \gg 0$ and
    choose $i \in \mathbb{N}$ such that $K^i$ is a compact linear operator on
    $L_p(\mathbb{X})$.  Fix $j \in \mathbb{N}$ with $0 \leq j \leq i-1$.
    By our assumptions on $K$, we know that $K^j h \gg 0$. Thus, theorem~\ref{t:lsr} applied to $K^i$
    with initial condition $K^j h$ yields
    \begin{equation*}
        \Big\{ \int (K^{in} K^j h)^p \, d\mu  \Big\}^{1/pn}
        = \Big\{ \int (K^{in+j} h)^p \, d\mu  \Big\}^{1/pn}
        \to \rho(K^i)
        \qquad (n \to \infty).
    \end{equation*}
    But $\rho(K^i) = \rho(K)^i$, so taking both sides to the power of $1/i$ yields
    \begin{equation*}
        \Big\{ \int (K^{in+j} h)^p \, d\mu  \Big\}^{1/(ipn)} \to \rho(K)
        \qquad (n \to \infty).
    \end{equation*}
    It follows that
    \begin{equation*}
        \Big\{ \int (K^{in+j} h)^p \, d\mu  \Big\}^{1/p(in+j)} \to
        \rho(K)
        \qquad (n \to \infty).
    \end{equation*}
    As $j$ is an arbitrary integer satisfying  $0 \leq j \leq i-1$,
    we conclude that \eqref{eq:lsr2} holds.
\end{proof}

The next lemma is a fixed point result which holds when $(\mathbb{X}, \mu)$ is a probability space.

\begin{lemma}
    \label{l:udfp}
    Let $\{g_n\}$ be a positive, monotone increasing sequence in $L_p(\mathbb{X})$.
    
    (1) If $\{g_n\}$ is bounded above by some $h$ in $L_p(\mathbb{X})$, then
    there exists a $g$ in $L_p(\mathbb{X})$ such that $\int g_n^p \: d\mu \to \int g^p \: d\mu$.
    
    (2) Moreover, let $g_n = T^n g_0$ for some continuous operator $T$
    mapping a subset of $L_p(\mathbb{X}, \mu)$ to itself. In this case, $g$ must be a fixed point of $T$.
\end{lemma}

\begin{proof}
    Regarding the first claim, note that since $\{g_n\}\subset L_p(\mathbb{X})$, we have that $\{g_n^p\}\subset L_1(\mathbb{X})$. Thus, by Beppo Levi's Monotone Convergence Theorem, $\int g_n^p \: d\mu \to \int g^p \: d\mu$ for some function $g^p \in L_p(\mathbb{X}, \mu).$ Then, since $\mathbb{X}$ is a finite measure space, applying Egorov's theorem shows that $\int |g_n - g|^p \: d\mu \to 0$. This $g$ must be the limit. This establishes the first part of the lemma.
    
    To see that $g$ is a fixed point of $T$, note that we have $|| g_n - g ||_p \to 0$ and
    hence, by continuity, $||T g_n - Tg||_p \to 0$.  But, by the definition of the sequence $\{g_n\}$,
    we also have $||T g_n - g||_p \to 0$.  Hence $Tg=g$.
    
    Note that in this proof, we rely heavily on the measure $\mu$ being finite. This is obviously satisfied as $\mu$ is a probability measure.
\end{proof}

\section{Proofs: Recursive Utility With Time Preference Shocks }

We now directly prove theorem \ref{BIG}. Our work draws heavily from the appendix of \cite{BorovickaStachurski2017}. Recall that we write the recursive utility operator as
$$Ag(x) = \Big\{ \xi(x) +  \beta\Big[Kg(x)\Big] ^{1/\theta} \Big\}^\theta$$
where $g \in L_p(\mathbb{X}, \mathcal{B}, \mu)$, $\theta \in \mathbb{R}$, and $\xi: \mathbb{X} \to \mathbb{R}$ is continuous and strictly positive.

Let $p$ represent the transition density for the exogenous state process $\{ X_t\} \subset \mathbb{X}$. By standard Markov process results, we may write the $i^{th}$ iteration as $p^i(x,y) = \int p(x,z) p^{i-1}(z,y) \: dz$ for all $x,y \in \mathbb{X}$. We assume that $p$ is irreducible in the sense that $p(x,y) > 0,\: \forall x,y \in \mathbb{X} $. We write
$$ k(x,y) =  \int  \exp[(1-\gamma)\kappa(x,y,\epsilon)] \nu(d\epsilon) p(x,y)  $$
We define a linear operator $K:L_p(\mathbb{X})_+ \to L_p(\mathbb{X})_+$ by
$$ Kg(x) = \int k(x,y) g(y) \: dy. $$
We also define $k^i$ as the $i^{th}$ iterate of $k$ such that $k^i(x,y) = \int k(x,z)k^{i-1}(z,y) dz.$ Thus, for all $x \in \mathbb{X}$ and $g \in L_p(\mathbb{X})$, we have
$$K^ig(x) = \int k^i(x,y) g(y) \: dy. $$
To see this, consider that
\begin{align*}
    K(Kg(x)) & = \int k(x,y) Kg(y) \: dy \\
    & = \int k(x,y) \int k(y,z) g(z) \: dz \: dy \\
    &= \mu(\mathbb{X}) \int k^2(x,y) g(y) \: dy \\
    & =  \int k^2(x,y) g(y) \: dy.
\end{align*}
From this a simple induction shows
$$K^i g(x) = \int k^i(x,y) g(y) \: dy.$$
The next three lemmas are $L_p$ versions of results found in \cite{BorovickaStachurski2017}.

\begin{lemma}
    \label{l:posd}
    The density $\pi$ is the unique stationary density for $p(x,\cdot)$ on $\mathbb{X}$.  In
    addition, $\pi$ is everywhere positive and continuous on $\mathbb{X}$.
\end{lemma}

\begin{proof}
    See \cite{BorovickaStachurski2017}, lemma B.1.
\end{proof}

\begin{lemma}
    \label{l:wpok}
    Regarding the operator $K$, the following statements are true:
    \begin{enumerate}
        \item $K$ is a bounded linear operator on $L_p(\mathbb{X}, \pi)$ that maps $L_p(\mathbb{X}, \pi)_+$ to itself. 
        \item $Kg \not= 0$ whenever $g \in L_p(\mathbb{X}, \pi)_+$ and $g \neq 0$.
        \item $Kg \gg 0$ whenever $g \in L_p(\mathbb{X}, \pi)$ and $g \gg 0$.
        \item For each $g\in L_p(\mathbb{X})_+$, $Kg$ is a continuous $L_p$ function.
    \end{enumerate}
\end{lemma}
\begin{proof}
    Regarding claim (a), $K$ is continuous and hence bounded by some constant
    $M$ on $\mathbb{X}$. Further, $\pi$ is positive and continuous on a compact set, and
    hence bounded below by some positive constant $\delta$.  This yields, for
    arbitrary $f \in L_p(\mathbb{X},\pi)$, and sufficiently large $N \in \mathbb{R}$
    \begin{align*}
        \label{eq:kib}
        |Kf(x)|^p &=
        \left| \int k(x, y) f(y) dy \right|^p \\
        &\leq M^p \Big( \int \frac{|f(y)|}{\pi(y)} \pi(y) dy\Big)^p \\
        &\leq  \frac{M^p}{\delta^p} \Big( \int |f(y)| \pi(y) dy\Big)^p \\
        &\leq  \frac{M^p N}{\delta^p} \Big( \int |f(y)|^p \pi(y) dy\Big)^{1/p} \\
        & = \frac{M^p N}{\delta^p} ||f||_p.
    \end{align*}
    The 2nd last inequality follows from the fact that as $(\mathbb{X}, \mu)$ is a finite measure space we get $L_p(\mathbb{X}, \mu) \subset L_1(\mathbb{X}, \mu). $ It follows directly that $K$ is a bounded linear operator on $L_p(\mathbb{X},\pi)_+$. Moreover, note that since $K$ is bounded, it must also be continuous. 

    Regarding claim (b), suppose that, to the contrary, we have
    $Kg = 0$ for some nonzero $g \in L_p(\mathbb{X})$.  Let $B = \{x \: : \:g(x) > 0\}$.
    Since $g$ is nonzero, we have $\pi(B) > 0$.  Since $Kg = 0$,
    it must be the case that $\int_B k(x, y) dy = 0$ for any $x \in \mathbb{X}$.
    But then $\int_B q(x, y) dy = 0$ for any $x \in \mathbb{X}$.
    A simple induction argument shows that this extends to the $n$-step
    kernels, so that, in particular, $\int_B p^\ell (x, y) dy
    = 0$ for all $x \in \mathbb{X}$.  The last equality contradicts $p^\ell > 0$, as guaranteed by
    irreducibility.

    Part (c) is immediate from $Kg(x) = \int g(y) k(x, y) dy$ and the definition of $k(x,y)$.
    
    To see part (d), fix $g \in L_p(\mathbb{X})_+, \: x \in \mathbb{X}$ and $x_n \to x$. Note that we have
    \begin{equation} k(x_n,y)g(y) \leq M \frac{g(y)}{\pi(y)}\pi(y) \leq \frac{M}{\delta}g(y)\pi(y) \label{bounding} \end{equation}
    Since $g \in L_p(\mathbb{X})_+$ and $\mathbb{X}$ has finite measure, it must be the case that $g$ is also in $L_1(\mathbb{X})$. Using \eqref{bounding}, we can apply the dominated convergence theorem to obtain
    $$ \lim_{n\to \infty} Kg(x_n) = \int \lim_{n \to \infty} k(x_n,y) g(y) \: dy = Kg(x). $$
    This shows that $Kg$ is continuous.
\end{proof}

\begin{definition} 
A linear operator $K: E \to E$ is called irreducible if $E$ and $\vec{0}$ are the only complementary invariant subspaces.
\end{definition}

\begin{lemma}
    \label{l:irr}
    The operator $K$ is irreducible and $K^2$ is compact. 
\end{lemma}

\begin{proof}
    To see that $K$ is irreducible, see \cite{BorovickaStachurski2017}, lemma B.3. 

    Regarding compactness, we must alter the argument somewhat. The theory of compact operators on an L-normed, Banach lattice implies that
    $K^2$ will be compact whenever $K$ is weakly compact, \footnote{To see this, I note that on an L-normed Banach lattice, if $T:E \to F$ is weakly compact, then $T(W)$ is precompact if $W$ is weakly-compact. I thank participants on math.stackexchange who showed this to me here:
    
    https://math.stackexchange.com/questions/3308257/product-of-two-weakly-compact-endomorphisms-is-compact } which requires that the
    image of the unit ball $B_1$ in $L_p(\mathbb{X},\pi)$ under $K$ is relatively compact
    in the weak topology.  To prove this it suffices to to show that, given
    $\epsilon > 0$, there exists a $\delta > 0$ such that $\int_A (K|f|)^p d
    \pi < \epsilon$ whenever $f \in B_1$ and $\pi(A) < \delta$.  This is true
    because $k$ is continuous and hence bounded on $\mathbb{X}$, yielding
    \begin{equation*}
        \int_A \Big(\int k(x, y) |f(y)| dy\Big)^p \pi(x) dx
        \leq \int_A \Big( \frac{M^pN}{\delta^p}||f||_p \Big) \pi(x) \: dx
        \leq \frac{M^pN}{\delta^p} ||f||_p \pi(A)
    \end{equation*}
    for constants $M,N$.  By taking $\pi(A) < \epsilon \delta^p / (M^pN ||f||_p ), $ we get the result.
\end{proof}

\begin{proposition}
    \label{p:lsr}
    $\Lambda_p$ is well defined and satisfies $\Lambda_p = \beta \, \rho(K)^{1/\theta}$.
\end{proposition}

\begin{proof}
    Since $K^i$ is compact for some $i$ and maps positive functions into
    positive functions (see lemmas~\ref{l:irr} and~\ref{l:wpok}), we can apply theorem~\ref{t:lsr2} to
    $\mathbb{1} \equiv 1$ to obtain $\rho(K) = \lim_{n \to \infty} \| K^n \mathbb{1} \|^{1/n}$.
    An inductive argument based on the \cite{BorovickaStachurski2017} consumption growth assumption shows that, for each $n$
    in $\mathbb{N}$, we have $K^n \mathbb{1} (x)
        = \mathbb{E}_x \left( C_n/C_0 \right)^{1-\gamma}$.
    Hence,
    \begin{equation}
        \label{eq:lim1}
        \| K^n \mathbb{1} \|_p^{1/n}
        =
        \left( \int\left\{
            \mathbb{E}_x \left( \frac{C_n}{C_0} \right)^{1-\gamma}
        \right\}^p \pi(dx) \right)^{1/np} 
        \\
        =
        \left( \mathbb{E}_\pi \left\{
            \mathbb{E}_{x} \left( \frac{C_n}{C_0} \right)^{1-\gamma}
        \right\}^p  \right)^{1/np} 
        \\
    \end{equation}
    Since $\rho(K) = \lim_{n \to \infty} \| K^n \mathbb{1} \|^{1/n}$, this yields
    \begin{equation*}
        \rho(K)
         =
        \lim_{n \to \infty} \,
        \left( \mathbb{E}_\pi \left\{
            \mathbb{E}_{x} \left( \frac{C_n}{C_0} \right)^{1-\gamma}
        \right\}^p  \right)^{1/np}
        =
        \mathcal{M}_{C,p}^{1 - \gamma}.
    \end{equation*} 
    Because $\theta := (1-\gamma)/(1 - 1/\psi)$, we now have
    \begin{equation*}
        \beta \rho(K)^{1/\theta}
        = \beta \mathcal{M}_{C,p}^{1 - 1/\psi}
        = \Lambda_p.
        \qedhere
    \end{equation*}
\end{proof}

\begin{theorem}
    \label{t:sri}
    The spectral radius $\rho(K)$ of $K$ is strictly positive.  Moreover, there
    exists an everywhere continuous eigenfunction $e$ of $K$ satisfying
    \begin{equation}
        \label{eq:pie}
        K e = \rho(K) e
        \quad \text{and} \quad
        e \gg 0.
    \end{equation}
\end{theorem}

\begin{proof}
    The
    irreducibility and compactness properties of $K$ obtained in
    lemma~\ref{l:irr} yield positivity of $\rho(K)$ and existence of the positive
    eigenfunction in equation \eqref{eq:pie}. This is by the De Pagter's theorem, and the Krein-Rutman theorem respectively.  Claim (d) of lemma~\ref{l:wpok} implies that $e$ is
    continuous, since $e \in L_p(\mathbb{X})$ and $e = (K e) / \rho(K)$.
\end{proof}

\begin{remark}
Let $\xi(x) \in C(\mathbb{X})$. As $\xi$ has compact support it is bounded. We also define $\xi > 0$. As such there exist constants $\xi_0, \xi_1 \in \mathbb{R}$ such that $\xi_0 < \xi(x)<\xi_1$ for all $x\in \mathbb{X}$. From this define $\varphi(t,x) = \Big\{ \xi(x) + \beta t^\frac{1}{\theta} \Big\}^\theta$, $\phi_0(t) = \Big\{ \xi_0 + \beta t^\frac{1}{\theta} \Big\}^\theta$, $\phi_1(t) = \Big\{ \xi_1 + \beta t^\frac{1}{\theta} \Big\}^\theta$.
\end{remark}

\begin{lemma}\label{lemma1} Let $A$ be defined as per the Schorfheide functional form.  Let $e$ be the Krein-Rutman eigenfunction of $K$. Let $\phi \in \Big\{ \phi_0, \phi_1 \Big\}$ refer to both $\phi_0$ and $\phi_1$. Suppose $\rho(K)$ is of the form that

\begin{align}
 a) \lim_{t\searrow 0} \frac{\phi(t)}{t}\rho(K) > 1 \: \: \text{ and }\: \: b) \lim_{t\nearrow \infty} \frac{\phi(t)}{t} \rho(K) < 1. \label{eq1}
\end{align}
Then we also have by boundedness and strict positivity of $\xi(x)$ that
 \begin{align}
a) \: \lim_{t\searrow 0} \frac{\varphi(t,x)}{t}\rho(K) > 1  \: \: \text{ and } \:\: b) \lim_{t\nearrow \infty} \frac{\varphi(t,x)}{t} \rho(K) < 1. \label{eq2}
 \end{align}
When this is the case, there exist positive constants $c_1<c_2$ such that

(1)  If $0 < c \leq c_1$ and $f = ce$, then there exists a $\delta_1 > 1$ such that $Af \geq \delta_1 f$

(2) If $c_2 \leq c < \infty$ and $f = ce$, then there exists a $\delta_2 < 1$ such that $Af \leq \delta_2 f$.

As a remark, note that the intuition here can be conceived as $\varphi$ being roughly `expansive' at first and then eventually `contractive' along eigenfunction paths. This intuitively will secure us a fixed point which is non-zero from any starting guess.
\end{lemma}

\begin{proof}[Proof] We first show that if condition \eqref{eq1} holds for $\phi_0$ and $\phi_1$ then condition \eqref{eq2} will also hold for $\varphi$. First, note that if there exists $\epsilon>0$ such that (a) of \eqref{eq1} holds for all $0 < t < \epsilon$, then by the joint continuity of $\varphi(t,x)$
$$ \frac{\varphi(t,x)}{t} \rho(K) \in \Big[ \min\Big\{\frac{\phi_0(t)}{t} \rho(K), \frac{\phi_1(t)}{t} \rho(K) \Big\} ,   \max \Big\{ \frac{\phi_0(t)}{t} \rho(K), \frac{\phi_1(t)}{t} \rho(K)\Big\}    \Big] . $$
Since the whole interval is greater than $1$, in view of expression \eqref{eq1} we get the result. Similarly, by part (b) of condition \eqref{eq1} there exists $M \in \mathbb{R}$ such that if $t > M$ then 
$$ \frac{\varphi(t,x)}{t} \rho(K) \in \Big[ \min\Big\{\frac{\phi_0(t)}{t} \rho(K), \frac{\phi_1(t)}{t} \rho(K) \Big\} ,   \max \Big\{ \frac{\phi_0(t)}{t} \rho(K), \frac{\phi_1(t)}{t} \rho(K)\Big\}    \Big] . $$ 
By the assumption that \eqref{eq1} holds for $\phi_0$ and $\phi_1$, the whole interval must now be less than one. This establishes \eqref{eq2}.

We now consider the first claim of the lemma. Let $e$ be the Perron-Frobenius (Krein-Rutman) eigenfunction of $K$. Let $\overline{e}$ and $\underline{e}$ be the maximum and minimum values of $e$ on $\mathbb{X}$ respectively. By \eqref{eq2} there exists a $\delta_1>1$ and $\epsilon > 0$ such that 
$$ \frac{\varphi(t,x)}{t}\rho(K) \geq \delta_1   $$
for all $x \in \mathbb{X}$ and $0<t<\epsilon.$
Now, choosing $c_1 \in \mathbb{R}$ such that $0<c_1 \rho(K) \overline{e} < \epsilon$ and $c \leq c_1$, we have $c \rho(K)e(x) < \epsilon$ for all $x \in \mathbb{X}$. Hence,
\begin{align*}
Ace(x) &= \varphi(c Ke(x),x) \\
&= \varphi(c \rho(K)e(x),x) \\
& = \frac{\varphi(c  \rho(K)e(x),x)}{c\rho(K)e(x)} c \rho(K)e(x) \\
& \geq \delta_1  c e(x)
\end{align*}
For the second statement, by \eqref{eq2} we may choose constants $\delta_2 < 1$ and $M < \infty$ such that 
$$ \frac{\varphi(t,x)}{t}\rho(K) \leq \delta_2  \: \: \: \text{ whenever } \: \: t > M $$
As such, choose $c_2 > \max \Big\{ \frac{M}{\rho(K)\underline{e} }, c_1 \Big\}$ and $c \geq c_2$. By definition of $\underline{e}$, we take $c \rho(K) e(x) \geq c_2 \rho(K) \underline{e} > M $ for all $x \in \mathbb{X}$. Hence
\begin{align*}
Ac\: e(x) & = \varphi(c \rho(K)e(x),x) \\
& = \frac{ \varphi(c  \rho(K) e(x), x  }{ c  \rho(K) e(x) }\rho(K)  c  e(x) \\
& \leq \delta_2  c  e(x).
\end{align*}
By construction of $0<c_1<c_2$ this completes the proof.
\end{proof}

\begin{lemma} \label{lemmab6} If the conditions from \eqref{eq2} hold and $A$ has a fixed point $g^* \in L_p(\mathbb{X})_+$ then there exist $f_1, f_2 \in L_p(\mathbb{X})_+$ such that 
$$f_1 \leq Ag, g^* \leq f_2, \: \: \:\: Af_1 \geq f_1 + \epsilon(f_2-f_1) \:\:\: \text{ and } \:\: Af_2 \leq f_2 - \epsilon(f_2-f_1)$$.
\end{lemma}
\begin{proof} Let $g \in L_p(\mathbb{X})_+$. Recall that $\xi(x) > 0$ for all $x\in\mathbb{X}$. Thus, since $Ag$ is continuous and $\mathbb{X}$ is compact, $Ag$ attains a finite maximum and strictly positive minimum. Similarly, the fixed point $g^* = Ag^*$ and Krein-Rutman eigenfunction $e(x)$ also attain finite maximum and strictly positive minimum.

From this, choose $a_1, a_2>0$ such that $0 \ll a_1 e \leq g^*,$ and $Ag \leq a_2 e$. If $a_1$ is small enough then lemma \ref{lemma1} implies $A(a_1e(x)) \geq \delta_1 a_1 e(x)$ for some $\delta_1 > 1$. If we then define $f_i := a_ie$, we have $Af_1 \geq \delta_1 a_1 e$. Since $\delta_1 > 1$, write $Af_1 \geq a_1 e + \epsilon_1(a_2-a_1)$ for small enough $\epsilon_1 >0$. From our definition of $f_1$ and $f_2$ we get the desired result that $Af_1 \geq f_1 + \epsilon_1(f_2 - f_1)$.

For the other inequality choose $a_2$ large enough that $f_2 = a_2 e$ and $Af_2 \leq \delta_2a_2 e$. Since $\delta_2 < 1$ then write $Af_2 \leq a_2 e - \epsilon_2(a_2-a_1)e = a_2 e - \epsilon_2(f_2-f_1).$

Choosing $\epsilon = \min \{ \epsilon_1, \epsilon_2 \}$ gives the overall result. 
\end{proof}

\begin{theorem}\label{thmb} If $\beta\rho(K)^\frac{1}{\theta}<1$, then $A$ is globally stable on $L_p(\mathbb{X})_+$.

\end{theorem}

\begin{proof} We first show that if $\beta\rho(K)^\frac{1}{\theta} < 1$ then the conditions in Lemma \eqref{eq2} hold. To see this, observe that
\begin{align}
\frac{\varphi(t,x)}{t} = \Big\{ \frac{\xi(x)}{t^\frac{1}{\theta}} + \beta \Big\} ^\theta \label{eq3}
\end{align}
where $\xi(x) \in [M_1,M_2]$ for some $M_1,M_2 \in \mathbb{R}$.

Consider the case where $\theta< 0$ with $\Lambda_p<1$. In this case we have $ \beta^\theta \rho(K)>1$ and, in addition, equation \eqref{eq3} increases to $\beta^\theta$ as $t \searrow 0$. Thus the first inequality of \eqref{eq2} holds. The second inequality then holds because $\varphi(t,x)/t \to 0$ as $t \to \infty $.

For the case where $\theta > 0$ we must have $\beta^\theta \rho(K) < 1$ and so $$ \Big\{ \frac{\xi(x)}{t^\frac{1}{\theta}} +\beta  \Big\} \nearrow \infty$$ as $t \searrow 0$. So the first inequality of \eqref{eq2} holds. The second inequality also holds because $ \beta^\theta\rho(K) < 1$ whilst $\varphi(t,x)/t \to \beta^\theta$ as $t \to \infty$. Thus choosing $t$ large enough will give the result for the second inequality.

This shows that the conditions in lemma \ref{lemma1} hold. To conclude the proof, note that for a fixed $x$, $\varphi(t,x)$ is either convex or concave in $t$, depending on $\theta$.

Suppose that $\varphi$ is concave in $t$. In this case, $A$ is isotone and concave in $g\in L_p{(\mathbb{X})}$ as a function from $L_p(\mathbb{X}) $ to $L_p(\mathbb{X})$. By lemma \ref{lemma1} then choose $c_1 < c_2$ such that $Ac_1e \geq c_1 e$ and $Ac_2e \leq c_2e$.

Applying \ref{du2} implies that $A$ has a fixed point $g^*\in L_p(\mathbb{X})_+$ which satisfies $c_1 e \leq g^* \leq c_2 e$. Since $e \gg 0$ and $c_1 > 0 $ we then get that $g^* \gg  0$. This gives us the uniqueness and existence of a fixed point.

To see global stability towards said fixed point, consider an arbitrary $g \in L_p(\mathbb{X})$. Choose $f_1, f_2$ as in lemma \ref{lemmab6}. This gives $f_1 \leq Ag \leq f_2$. Then by \ref{du2} we have that every element of $[f_1,f_2]$ converges to $g^*$ under $A$. In particular, $A^n(Ag) \to g^*$ in norm as $n \to \infty$ by virtue of our definition of $f_i$. But then $A^ng \to g^*$ also holds and so $A$ is stable on $L_p(\mathbb{X})$.

The convex case is largely the same.
\end{proof}

\hfill

\begin{proposition}[Necessity]\label{necessity} If $A$ has a nonzero fixed point in $L_p(\mathbb{X})_+$, then $\beta \rho(K)^\frac{1}{\theta}<1$.
\end{proposition}

\begin{proof}
 Recall that $K$ is a linear operator on a Banach space. As such, let $K^*$ be the adjoint operator. Since $K$ is irreducible and $K^2$ is compact note that by De Pagter's theorem,  $\rho(K) >0.$ Thus, by the Krein-Rutman and Riesz representation theorems, for $q$ satisfying $ \frac{1}{q} + \frac{1}{p}= 1$ we get the existence of $e^* \in L_q(\mathbb{X})$ such that
$$ e^* \gg 0 \text{ and } \: K^*e^* = \rho(K)e^*. $$

Before proceeding, it is helpful to note that if $f^* \in L_q(\mathbb{X})_+$ and $g \in L_p(\mathbb{X})_+$ then $\int f^*(x) g(x) \: d\pi < \infty$ is well defined. This can be seen by observing that if $\int |f^{*}(x)|^q \: d\pi < \infty$ and $\int |g(x)|^p \: d\pi < \infty$ then the embedding $L_p,L_q \subset L_1$ on a finite measure space implies $\int f^*(x) \: d\pi \int g(x) \: d\pi < \infty$. Hence, by Cauchy-Schwarz
\begin{align*}
\int f^*(x) g(x) \: d\pi & \leq \int f^*(x) \: d\pi \int g(x) \: d\pi < \infty.
\end{align*}

With this in mind, define $g$ to be a nonzero fixed point of $A$ in $L_p(\mathbb{X})_+$. We now prove the proposition for $\theta < 0$. In this case we have $\varphi(t,x) <  \beta^\theta t$ whenever $t > 0$ due to the fact that $\xi \gg 0$. By assumption, we then also have that $Kg \gg 0$, so $g(x) = Ag(x) = \varphi(Kg(x)) < \beta^\theta Kg(x).$ Since $e^* \gg 0$ it follows that $ \int e^*(x)(\beta^\theta Kg(x)-g(x))\: d\pi> 0$. Using the definition of the adjoint then shows that
$$ \rho(K) \int e^*(x) g(x) \: d\pi = \int\beta^\theta  K^*e^*(x) g(x) \: d\pi = \int \beta^\theta e^*(x) Kg(x) \: d\pi.$$
Combining these two inequalities, it must be the case that $\beta^\theta \rho(K) \int e^*(x) g(x) \: d\pi > \int e^*(x) g(x) \: d\pi$. Since $\theta < 0$, this shows that $\Lambda_p = \beta\rho(K)^{1/\theta} < 1$.

For the case where $\theta > 0$, note that $\varphi(t,x) > \beta^\theta t$ whenever $t > 0$. As we again know that $Kg \gg 0$ it must be the case that $g(x) = Ag(x) = \varphi(Kg(x)) > \beta^\theta Kg(x).$ By a symmetric argument to above, 
$$\beta^\theta \rho(K) \int e^*(x)g(x) \: d\pi = \beta^\theta \int K^*e^*(x)g(x) \: d\pi = \beta^\theta\int e^*(x)Kg(x) \: d\pi < \int e^*(x) g(x) \: d\pi.$$ 
Hence $\beta^\theta \rho(K) <1$, and so $\beta \rho(K)^\frac{1}{\theta} < 1.$
\end{proof}

\begin{proof}[Proof of theorem \ref{BIG}]
We first note that that (e) $\implies$ (d). This is due $K$ being a bounded linear operator on $L_p(\mathbb{X})$ and $\varphi$ being jointly continuous on $\mathbb{R}_+^2$. Hence, it follows that $A$ is continuous on $L_p(\mathbb{X})_+$, and so any limit of a sequence of iterates $\{ A^ng \}_{n \geq 1} $ of $A$ is a fixed point of $A$. As the limit is unique from any starting point, the fixed point is unique.

Moreover, (d) $\implies$ (c) by taking $g$ equal to the fixed point. Furthermore, (c) $\implies $ (b). This is again by continuity of $A$ on $L_p(\mathbb{X})_+$, meaning that any limit of a sequence $\{ A^ng \}_{n \geq 1} $ of $A$ is a fixed point of $A$. 

The implication (b) $\implies$ (a) is due to proposition \ref{necessity}. Finally (a) $\implies $ (e) by theorem \ref{thmb} and proposition \ref{p:lsr}.
\end{proof}

\section{Proofs: Recursive Utility with Narrow Framing}

This section proves proposition \ref{narrow}. The proofs here have direct analogues from the previous section.

\begin{remark}
Let $b(x) \in C(\mathbb{X})$. As $b$ has compact support it is bounded. We also define $b > 0$. As such there exist constants $\xi_0, \xi_1 \in \mathbb{R}$ such that $b_0 < b(x)<b_1$ for all $x\in \mathbb{X}$.
\end{remark}

\begin{lemma}\label{c1}
 Let $A$ be the recursive utility with narrow framing operator.  Let $e$ be the Krein-Rutman eigenfunction of $K$. Let $\phi(t)$ be defined as earlier and redefine $\varphi(t,x) = \Big\{ 1 -\beta + \beta \Big(  t + b(x) \Big)^\frac{1}{\theta} \Big\}^\theta$ . Suppose $\rho(K)$ and $\phi(t)$ satisfy
\begin{align}
 a) \lim_{t\searrow 0} \frac{\phi(t)}{t}\rho(K) > 1 \: \: \text{ and }\: \: b) \lim_{t\nearrow \infty} \frac{\phi(t)}{t} \rho(K) < 1. \label{ceq1}
\end{align}
Then we also have by boundedness and strict positivity of $b(x)$ that $\varphi(t,x)$ satisfies
 \begin{align}
a) \: \lim_{t\searrow 0} \frac{\varphi(t,x)}{t}\rho(K) > 1  \: \: \text{ and } \:\: b) \lim_{t\nearrow \infty} \frac{\varphi(t,x)}{t} \rho(K) < 1. \label{ceq2}
 \end{align}
When this is the case there exist positive constants $c_1<c_2$ such that

1)  If $0 < c \leq c_1$ and $f = ce$, then there exists a $\delta_1 > 1$ such that $Af \geq \delta_1 f$

2) If $c_2 \leq c < \infty$ and $f = ce$, then there exists a $\delta_2 < 1$ such that $Af \leq \delta_2 f$.
\end{lemma}
\begin{proof}
We first show that condition \textit{a)} translates from \eqref{ceq1} to \eqref{ceq2}. In this respect note that for any value of $\theta \in \mathbb{R}_{\neq 0}$, if $t$ is small enough to satisfy \textit{a)}, then
\begin{align*}
    1 & < \frac{\phi(t)}{t} \rho(K) \\
    & =\rho(K) \Big[ \frac{1-\beta}{t^\frac{1}{\theta} } + \beta \Big( \frac{t}{t} \Big) ^ \frac{1}{\theta} \Big]^\theta \\
    & \leq  \rho(K) \Big[ \frac{1-\beta}{t^\frac{1}{\theta} } + \beta \Big( \frac{t + b(x)}{t} \Big) ^ \frac{1}{\theta} \Big]^\theta \\
    & = \frac{\varphi(t,x)}{t}\rho(K)
\end{align*}
by positivity of $b(x)$.

We now consider part \textit{b)} of the translation. This is straightforward using the observation that since $b(x)$ is bounded for all $x \in \mathbb{X}$ we have
\begin{align*}
\lim_{t\nearrow \infty} \frac{\phi(t)}{t} & = \lim_{t \nearrow \infty} \Big[ \frac{1-\beta}{t^\frac{1}{\theta} } + \beta \Big( \frac{t}{t} \Big) ^ \frac{1}{\theta} \Big]^\theta \\
&= \lim_{t \nearrow \infty} \Big[ \frac{1-\beta}{t^\frac{1}{\theta} } + \beta \Big( \frac{t + b(x)}{t} \Big) ^ \frac{1}{\theta} \Big]^\theta \\ 
& = \lim_{t\nearrow \infty} \frac{\varphi(t,x)}{t}.
\end{align*}
The rest of the proof then follows identically from \ref{lemma1}.
\end{proof}

\begin{proposition}
 If the conditions from equation \eqref{ceq2} in lemma \ref{c1} hold and $A$ has a fixed point $g^* \in L_p(\mathbb{X})$ then there exist $f_1, f_2 \in L_p(\mathbb{X})$ such that 
$$f_1 \leq Ag, g^* \leq f_2, \: \: \:\: Af_1 \geq f_1 + \epsilon(f_2-f_1) \:\:\: \text{ and } \:\: Af_2 \leq f_2 - \epsilon(f_2-f_1)$$.
\end{proposition}
\begin{proof}
This follows by a similar argument to lemma \ref{lemmab6} replacing $\xi(x)$ with $b(x)$.
\end{proof}

\begin{proposition}[Sufficiency]
 If $\beta \rho(K)^\frac{1}{\theta}<1$, then $A$ is globally stable on $L_p(\mathbb{X})$.
\end{proposition}
\begin{proof} This proof proceeds very similarly to \ref{thmb}.  We first show that if $\beta \rho(K)^\frac{1}{\theta} < 1$ then the conditions in Lemma \eqref{ceq2} hold. To see this, observe that
\begin{align}
\frac{\varphi(t,x)}{t} = \Big\{ \frac{1-\beta}{t^\frac{1}{\theta}} + \beta \Big(\frac{t + b(x) }{t} \Big)^\frac{1}{\theta}  \Big\} ^\theta \label{ceq3}
\end{align}
where $b(x) \in [M_1,M_2]$ for some $M_1,M_2 \in \mathbb{R}$.

Consider the case where $\theta< 0$. In this case we have $\beta^\theta \rho(K)>1$ and so equation \eqref{ceq3} grows arbitrarily large as $t \searrow 0$. Thus the first inequality of \eqref{ceq2} holds. The second inequality then holds because $\varphi(t,x)/t \to 0$ as $t \to \infty $.

For the case where $\theta > 0$ we must have $\beta^\theta\rho(K) < 1$ and so $$ \Big\{ \frac{1-\beta}{t^\frac{1}{\theta}} + \beta\Big(\frac{t + b(x) }{t} \Big)^\frac{1}{\theta}  \Big\} \nearrow \infty$$ as $t \searrow 0$. So the first inequality of \eqref{ceq2} holds. The second inequality also holds because $\beta^\theta \rho(K) < 1$ whilst $\varphi(t,x)/t \to \beta^\theta$ as $t \to \infty$. Thus choosing $t$ large enough will give the result for the second inequality.

This shows that the conditions in lemma \ref{c1} hold. To conclude the proof, note that for a fixed $x$, $\varphi(t,x)$ is either convex or concave in $t$, depending on $\theta$. The rest of the proof follows in exactly the same manner as \ref{thmb}.

\end{proof}

\section{ Proofs: Recursive Utility on an Unbounded State Space }

In this section we prove theorem \ref{unbounded}. The following lemmas generalise results found in the online appendix of \cite{BorovickaStachurski2017}.

\begin{lemma}\label{2point2}
Let $\{T_n\}$ and $T$ be bounded linear operators on $L_p(\mathbb{X},\pi)$ such that $0 \leq T_n \leq T_{n+1} \leq T$ for all $n \in \mathbb{N}$. If $\int |T_nf - Tf|^p d\pi \to 0$ as $n\to \infty$ for each $f$ in the positive cone $L_p(\mathbb{X})_+$ and $T^i$ is compact for some $i \in \mathbb{N},$ then $\rho(T_n) \nearrow \rho(T).$
\end{lemma}
\begin{proof}
The proof follows identically to lemma 2.2 of the online appendix of \cite{BorovickaStachurski2017}, replacing the $L_1$ norm with the $L_p$ norm. In particular, the spectral continuity result of \cite{schep1980} will apply to all $L_p$ spaces.
\end{proof}

\begin{lemma}\label{2point3}
Let (E,d) be a metric space and let $T$ and $\{T_m\}_{m \in \mathbb{N}}$ be operators on $E$ with the property that $T_mu \to Tu$ in norm for all $u \in E$. Let $\bar{u}_m$ be a fixed point of $T_m$ for each $m$ and suppose that $\bar{u}_m \to \bar{u}$ for some $\bar{u} \in E$. If $T$ is continuous on $E$ and the maps $\{T_m\}$ are uniformly Lipschitz continuous, then $\bar{u}$ is a fixed point of $T$.
\end{lemma}
\begin{proof}
See \cite{BorovickaStachurski2017} online appendix, lemma 2.3.
\end{proof}

We now turn to more direct results used in the proof of theorem \ref{unbounded}. This proof relies on a limiting argument based on approximating $\mathbb{X}$ with compact sets. The proofs build upon those seen in the online appendix of \cite{BorovickaStachurski2017}.

Let $\{F_m\}_{m\in\mathbb{N}}$ be an increasing sequence of compact sets such that $F_m \subset F_{m+1}$ for all $m\in\mathbb{N}$. Since $\mathbb{X}$ is $\sigma-$finite, we set $\bigcup_{m\in\mathbb{N}} F_m = \mathbb{X}.$ Let $K_m$ be the operator on $L_p(\mathbb{X},\pi)$ defined by 
$$ K_mg(x) = \mathbb{1}_{x\in F_m} \int_{F_m} k(x,y) g(y) \: dy  $$
Note that $K_m$ is also a positive linear operator and $0 \leq K_m \leq K_{m+1}$ for all $m \in \mathbb{N}$. Then $K_m$ is a bounded linear operator on $L_p(\mathbb{X},\pi)$.

\begin{lemma}\label{2point5}
If $f \in L_p(\mathbb{X},\pi)_+$, then $||K_mf - Kf|| \to 0$ as $m \to \infty$.
\end{lemma}
\begin{proof}
Fix $f \in L_p(\mathbb{X},\pi)_+$. For any $m \in \mathbb{N},$ we have 
$$ ||K_mf - Kf||^p \leq \int \left( \int k(x,y) \left(1 - \mathbb{1}_{F_m}(x)\mathbb{1}_{F_m}(y)\right) f(y) \: dy\right)^p\: d\pi(x).  $$
Since $K$ is a bounded, linear operator, the integral on the right hand side is finite. Since we are on a finite measure space, Egorov's theorem means that it suffices to show that the integrand converges pointwise to $0$. This follows immediately from the definition of $\{F_m\}$.
\end{proof}

Given $g:F_m \to \mathbb{R}$, as per \cite{BorovickaStachurski2017}, define its \textit{extension} $e_mg$ to $\mathbb{X}$ as the function equal to $g$ on $F_m$ and zero on $F_m^c$. Given $g:\mathbb{X} \to \mathbb{R},$ its \textit{restriction} $c_mg$ to $F_m$ is defined as the function $c_mg$ equal to $g$ on $F_m$. In addition, let $\bar{K}$ be the restriction of $K_m$ to real functions on $F_m$. That is,
$$ \bar{K}_m g(x) = \int_{F_m} k(x,y) g(y) dy . $$
We regard $\bar{K}_m$ as a mapping on $L_p(F_m,\bar{\pi})$, where $\bar{\pi} := c_m \pi.$ Note that
\begin{equation}
A_m = e_m \bar{A}_mc_m \label{onesy}
\end{equation}
on $L_p(\mathbb{X})_+$, where $A_m = \varphi \circ K_m$ and $\bar{A}_m := \varphi \circ \bar{K}_m$. The latter is a self-mapping on the positive cone $L_p(F_m, \bar{\pi}_m)_+.$

\begin{lemma}\label{2point6}
If $g \in L_p(F_m, \bar{\pi}_m)_+$ is a fixed point of $\bar{A}_m$, then $e_mg$ is a fixed point of $A_m$.
\end{lemma}
\begin{proof}
For $g \in L_p(F_m, \bar{\pi})_+$ we have $A_m e_m g = e_m \bar{A}_m c_m e_m g = e_m \bar{A}_m g = e_m g$.
\end{proof}

\begin{lemma}\label{2point7}
For all $m \in \mathbb{N}$, we have $||\bar{K}_m|| = ||K_m||$.
\end{lemma}
\begin{proof}
Fix $f \in L_p(\mathbb{X}, \pi)$ with $||f|| \leq 1$. Let $\bar{f}$ be the restriction of $f$ to $F_m$. Note that,
$$ ||\bar{f}||^p = \int |\bar{f}|^p \bar{\pi}(x) dx \leq ||f||^p \leq 1.  $$
We have
$$ ||\bar{K}\bar{f} ||^p = \int_{F_m} \left| \int_{F_m} k(x,y)f(x) \: dy \right|^p \pi(x) dx = \int \left|K_mf(x)\right|^p\pi(x) dx = ||K_mf||^p. $$
Thus, by the definition of the operator norm we have that
$$ ||K_mf|| = ||\bar{K}_m \bar{f}|| \leq ||\bar{K}_m|| $$
and then by taking the supremum over $\{K_mf \: : \: ||f||\leq 1\}$ on the left hand side we get $||K_m|| \leq ||\bar{K}_m||$.

To see the reverse inequality holds, fix $\bar{f} \in L_p(F_m, \bar{\pi})$ with $||\bar{f}|| \leq 1.$ Let $f \in L_p(\mathbb{X},\pi)$ be defined by $f = \bar{f}$ on $F_m$ and $f = 0$ elsewhere. Note that
$$ ||f||^p = \int |f|^p \pi(x) \: dx  = \int |\bar{f}|^p \bar{\pi}(x) dx = ||\bar{f}||^p \leq 1. $$
By an identical argument to above this gives $||\bar{K} \bar{f} || = ||K_m f||.$ It follows that $ ||\bar{K}_m \bar{f}|| \leq ||K_m||,$ and taking the supremum on the left over all such $\bar{f}$ yields $||\bar{K}_m|| \leq ||K_m||$.
\end{proof}

\begin{lemma}\label{2point8}
If $\rho(K) > 1/\beta^\theta$, then there exists an $M \in \mathbb{N}$ such that $\rho(\bar{K}_m) > 1/\beta^\theta$ whenever $m \geq M$.
\end{lemma}
\begin{proof}
In view of lemma \ref{2point7} and the definition of the spectral radius, it suffices to prove that $\rho(K_m) > 1$, for sufficiently large $m$. This will be true if $\rho(K_m) \to \rho(K)$, which, by lemma \ref{2point2}, will hold if (a) $K^i$ is compact for some $i \in \mathbb{N}$, (b) $0 \leq K_m \leq K_{m+1} \leq K$ for all $m$ and (c) $K_mf \to Kf$ in norm for each $f$ in $L_p(\mathbb{x}, \pi)_+$. We already have (a) by eventual compactness and (b) is true by construction. Finally, (c) holds by lemma \ref{2point5}.
\end{proof}

\begin{lemma}\label{2point9}
Under the conditions of theorem \ref{unbounded}, $\Lambda_p$ is well defined and satisfies $\Lambda_p = \beta \rho(K)^{1/\theta}.$
\end{lemma}
\begin{proof}
Proof is identical to the compact state space case in proposition \ref{p:lsr}.
\end{proof}

\begin{lemma}\label{2point10}
If $\theta < 0$ and $\Lambda_p <1$, then there exists an $M \in \mathbb{N}$ such that, for all $m \geq M,$ the operator $A_m$ has a nonzero fixed point $g_m \in L_p(\mathbb{X}, \pi)_+$, and $g_m \leq g_{m+1}$ for all such $m\in \mathbb{N}$.
\end{lemma}
\begin{proof}
If $\theta < 0$ and $\Lambda_p<1$, by proposition \ref{2point9}, we have $\rho(K) > 1/\beta^\theta$. Now, let $M$ be as in lemma \ref{2point8} and take $m \geq M$. Observe that $\bar{A}_m$ has a unique nonzero fixed point $\bar{g}_m$ in $L_p(\mathbb{X}, \pi)$, since $F_m$ is compact. It then follows from equation \eqref{onesy} that 
$$ A_me_m\bar{g}_m = e_m\bar{A}_mc_me_m\bar{g}_m = e_m\bar{g}_m  $$
and so
$$g_m:=e_m\bar{g}_m $$
is a fixed point of $A_m$. Since $\bar{g}_m$ is nonzero on $F_m$, the function $g_m$ is nonzero on $\mathbb{X}$.

It remains to prove that $g_m \leq g_{m+1}$ for all $m \geq M$. As such, choose some $m \geq M$ and note that since $K_m \leq K_{m+1}$ on $L_p(\mathbb{X}, \pi)$ and $\varphi$ is increasing, we have $A_{m+1}g_m \geq A_mg_m = g_m$. Using isotonicity of $A_{m+1}$ and iterating forward yields $A^n_{m+1}g_m \geq g_m$ for all $n \in \mathbb{N}$. Moreover, since $g_m$ is nonzero on $F_m$ and hence $F_{m+1},$ the convergence result of theorem \ref{BIG} applied to the compact set $F_{m+1}$ implies that $A_{m+1}^ng_m \to g_{m+1}$ uniformly. Hence $g_{m+1} \geq g_m$, as was to be shown.
\end{proof}

\begin{lemma}\label{2point11}
If $\theta < 0$, then the family $\{A_n\}$ is uniformly Lipschitz continuous on $L_p(\mathbb{X}, \pi)_+$.
\end{lemma}
\begin{proof}
When $\theta < 0$, the scalar map $\varphi$ is Lipschitz with Lipschitz constant 1. Hence, for arbitrary $m \in \mathbb{N}$ and $f,g \in L_p(\mathbb{X},\pi)_+$ we have
$$ |A_mf - A_mg| \leq |K_mf - K_mg| = |K_m(f-g)| \leq K_m|f-g| \leq K|f-g|   $$
from monotonicity of Lebesgue integration we then get
$$ \int |A_mf-A_mg|^p d\pi \leq \int \left(K|f-g|\right)^pd\pi $$
which in turn implies
$$||A_mf - A_mg|| \leq ||K|| \cdot ||f-g||  $$
\end{proof}
This brings us to the final proof of theorem \ref{unbounded}.

\begin{proof}[Proof of Theorem \ref{unbounded}.]

We first show that (a) $\iff$ (b) under the assumptions claimed. We begin with (a) $\implies$ (b). When $\theta > 0$, the proof follows directly from the proof of the compact case, as the proof does not rely on compactness. 

Thus, suppose that $\theta < 0$. Observe that by lemma \ref{2point5}, for $f \in L_p(\mathbb{X},\pi)_+$ we have $K_mf \to Kf$ as $m \to \infty$. Since $\varphi$ is Lipschitz continuous of order 1 when $\theta < 0$, we get immediately that $A_mf \to Af$ as $m \to \infty.$ By lemma \ref{2point10}, there exists an $M \in \mathbb{N}$ such that for all $m \geq M$, the operator $A_m$ has a nonzero fixed point $g_m \in L_p(\mathbb{X},\pi)_+$, and $g_m \leq g_{m+1}$ for all such $m$. Since $\varphi$ is bounded above by $(1-\beta)^\theta$ when $\theta < 0$ it must be that $g_m \leq g_{m+1} \leq (1-\beta)^\theta$ for all $m$.

Note, any order bounded monotone sequence in $L_1(\mathbb{X}, \pi)$ converges to an element of that set. Denote the limit by $g$. Then since $\int |g_m(x) - g(x)| d\pi \to 0$, observing that we are on a finite measure space and applying Egorov's theorem shows that $\int |g_m(x) - g(x)|^p d\pi \to 0$. Hence, we get that $g$ is also the $L_p(\mathbb{X}, \pi)$ limit. In view of lemma \ref{2point3}, this $g$ will be a fixed point of $A$ whenever $A$ is continuous and $\{A_m\}$ is uniformly Lipschitz continuous. Continuity of $A$ is immediate from the properties of $K$ and $\varphi,$ while uniform Lipschitz continuity of $\{A_m\}$ follows from lemma \ref{2point11}. This shows that (a) $\implies$ (b).

We now show (b) $\implies$ (a). In this case the exact same proof as used when $\mathbb{X}$ was compact can be used. In proposition \ref{necessity} compactness was only used to ensure that $K^i$ was compact on $L_p(\mathbb{X}, \pi)$ for some $i \in \mathbb{N}$. In the unbounded setting, this condition still holds by eventual compactness. This shows that (b) $\implies$ (a).

Finally, the equivalence between (b) and (c) holds by standard arguments. The fact that (b) $\implies$ (c) follows by choosing $g$ to be the fixed point in the statement of (c). To see that (c) $\implies$ (b) note that $g^* = \lim_{n\to \infty}A^ng$ is a fixed point of $A$ by continuity of $A$.

\end{proof}

\section{Proofs from Chapter 4}

\begin{proof}[Proof of Theorem \ref{approxthm}]

From the outset, we take $x_2 \in [0, M]$.

Now note that for some $A_\sigma> 0 $, $k(x,y)$ must satisfy
\begin{align*}
    k(x,y) &=  \exp\Big[(1-\gamma)(\mu_c + x_1) + (1-\gamma)^2 \frac{\sqrt{ x_2+\epsilon} }{ 2 }\Big] \cdot q(x,y)\\
    &\leq A_\sigma \frac{ \exp\Big[(1-\gamma)(\mu_c + x_1) + (1-\gamma)^2 \frac{\sqrt{ x_2+\epsilon} }{ 2 }\Big] }{\sqrt{2\pi (x_2+\epsilon)} }\cdot \exp\Big[\frac{-(y_1-\rho x_1)^2 }{ 2 (x_2+\epsilon)}\Big].
\end{align*}
Since $x_2$ is bounded, we can take sufficiently large $B,C,D > 0$ such that for some sufficiently large ball around the origin $B_{r>M}(0) = \Omega \subset \mathbb{R}^2 \times \mathbb{R}^2$, we have
\begin{align*}
    \int |k(x,y)|^2 d(\pi\times \pi) &\leq D \int \exp[Bx_1 + C] d(\pi\times \pi)\\
    & = D \Big( \int_\Omega  \exp[Bx_1 + C] d(\pi\times \pi) + \int_{\Omega^c}  \exp[Bx_1 + C] d(\pi\times \pi)  \Big)\\
    & \leq D\Big( Q_M  + R_N\int_{\Omega^c} \exp[Bx_1 + C]  \exp\Big[ \frac{-x_1^2}{2M^2}  \Big] dx_1  \Big)\\
    &< \infty
\end{align*}
for some $Q_M, R_N >0$. In particular, the second last line follows from the fact that the tail density of $x_1$ is weakly dominated by a normal distribution with variance $M^2$. We take a large ball around the origin so that we need only integrate around this tail density in the latter term.
\end{proof}

\hfill

\begin{proof}[Proof of Theorem \ref{HilbertSchmidt2}]

The Schwartz kernel is given by
\begin{align*}
m(x,y) &= \int \exp\left[ (1 - \gamma) ( \mu_c + x_3 + (\phi_c \bar{\sigma} e^{x_1})\epsilon  \right]  \nu(d\epsilon) q(x,y) \\
& = q(x,y) \exp\Big[(1-\gamma)(\mu_c + x_3) \Big] \cdot  \exp\Big[ \frac{(\phi_\epsilon \bar{\sigma} e^{x_1})^2}{2} \Big]
\end{align*}
 where by assumption $x_1$ is bounded.

 Using proposition \ref{L2}, it suffices to verify that $m(x,y) \in L_2(\mathbb{R}^3 \times \mathbb{R}^3)$. 

Note that $q = q((x_1, x_2, x_3), (y_1,y_2, y_3)) \leq A \: q( (\cdot , \cdot, x_3), (\cdot, \cdot, y_3))$\footnote{That is, consider only the `slice' density over $x_3$, and allow all other arguments to take any value.} for some $A >0$. Thus
\begin{align*}
     q(x,y) & \leq \frac{A}{R} \exp\Big[ \frac{-(y_3 - \rho x_3)^2 }{ B e^{ 2x_2 } } \Big] \mathbb{1}_{-M \leq x_2 \leq M} \\
     & \leq \frac{A}{R} \exp \Big[ \frac{-(y_3 - \rho x_3)^2 }{B \cdot e^{2M}}\Big]
\end{align*}
for $B =2 \sqrt{1-\rho^2} \phi_z \bar{\sigma}$, and $R$ the constant of normalisation. Thus, for $Q_M>0$ sufficiently large we may bound the Schwartz kernel by
\begin{align*}
    m(x,y) &\leq \frac{A}{R} \exp \Big[ \frac{-(y_3 - \rho x_3)^2 }{B \cdot e^M}\Big] \exp\Big[(1-\gamma)(\mu_c + x_3) \Big] \cdot  \exp\Big[ \frac{(\phi_\epsilon \bar{\sigma} e^{x_1})^2}{2} \Big]\\
    &\leq Q_M \cdot \exp\Big[ (1-\gamma) x_3 - \frac{(y_3 - \rho x_3)^2}{B e^{2M}}\Big] \\
    &\leq Q_M \cdot \exp\Big[ (1-\gamma) x_3 \Big]
\end{align*}
Thus, note that for some sufficiently large ball around the origin $B_{r>2M}(0) = \Omega \subset \mathbb{R}^3 \times \mathbb{R}^3$, and sufficiently large $N,C > 0$
\begin{align*}
 \int |m(x,y)|^2 d\pi &\leq Q_M^2 \int e^{2(1-\gamma)x_3} d \pi \\
 &= Q_M^2 \Big( \int_\Omega e^{2(1-\gamma)x_3} d\pi + \int_{\Omega^c} e^{2(1-\gamma)x_3} d\pi  \Big) \\
 & \leq N + Q_M^2 \int_{\Omega^c} e^{2(1-\gamma)x_3} d\pi \\
 & \leq N + Q_M^2C \int_{\Omega^c} e^{2(1-\gamma)x_3}e^{\frac{-x_3^2 \varphi}{ e^M} } dx_3 \\
 & < \infty
 \end{align*}
where $\varphi$ is a constant of normalisation.\footnote{ The second last line of the argument follows by noting that the tail distribution of $x_3$ must be sub-Gaussian (see Lemma 2.1.1 of \cite{HDP}), and that the density with respect to $(x_1,x_2,y_1,y_2,y_3)$ decays to be uniformly bounded by 1 for large values.} That is, $m(x,y) \in L_2(\mathbb{R}^3 \times \mathbb{R}^3).$
\end{proof}

\bibliographystyle{abbrvnat}
\bibliography{main}

\end{document}